\documentclass{article}

\usepackage[final]{neurips_2022}

\usepackage[utf8]{inputenc} 
\usepackage[T1]{fontenc}    
\usepackage{hyperref}       
\usepackage{url}            
\usepackage{booktabs}       
\usepackage{amsfonts}       
\usepackage{nicefrac}       
\usepackage{microtype}      
\usepackage{xcolor}         

\usepackage{mathtools}  
\usepackage{caption}
\usepackage{subcaption}

\usepackage{floatflt}
\usepackage{listings}

\usepackage{tikz}
\usepackage{layouts}

\usepackage{xcolor}

\definecolor{codegreen}{rgb}{0,0.6,0}
\definecolor{codegray}{rgb}{0.5,0.5,0.5}
\definecolor{codepurple}{rgb}{0.58,0,0.82}
\definecolor{backcolour}{rgb}{0.95,0.95,0.92}

\lstdefinestyle{mystyle}{
    backgroundcolor=\color{backcolour},   
    commentstyle=\color{codegreen},
    keywordstyle=\color{magenta},
    numberstyle=\tiny\color{codegray},
    stringstyle=\color{codepurple},
    basicstyle=\ttfamily\footnotesize,
    breakatwhitespace=false,
    breaklines=true,                 
    captionpos=b,                    
    keepspaces=true,                 
    numbers=left,                    
    numbersep=5pt,                  
    showspaces=false,                
    showstringspaces=false,
    showtabs=false,                  
    tabsize=2
}

\newcommand{\ts}{\textsuperscript}

\title{Differentiable Programming of Chemical Reaction Networks}

\author{%
  Alexander Mordvintsev \\
  Google Research\\
  \texttt{moralex@google.com} \\
  \And
  Ettore Randazzo \thanks{Equal contribution.} \\
  Google Research \\
  \texttt{etr@google.com} \\
  \AND
  Eyvind Niklasson \footnotemark[1] \\
  Google Research\\
  \texttt{eyvind@google.com} \\
}

\begin{document}

\maketitle

\begin{abstract}
  We present a differentiable formulation of abstract chemical reaction networks (CRNs) that can be trained to solve a variety of computational tasks. Chemical reaction networks are one of the most fundamental computational substrates used by nature. We study well-mixed single-chamber systems, as well as systems with multiple chambers separated by membranes, under mass-action kinetics. We demonstrate that differentiable optimisation, combined with proper regularisation, can discover non-trivial sparse reaction networks that can implement various sorts of oscillators and other chemical computing devices.
\end{abstract}

\section{Introduction}

Computation and information processing, implemented using different physical substrates and at different scale, are ubiquitous in nature and in technology.  The most effective computational devices rely on fine and persistent physical structures. Examples are natural neural networks and human-made electronic, mechanical, or hydraulic computers.

There is another very important class of computational networks, which are responsible for decision making at scales from individual cells to societies, that are much less demanding of the precise spatial structure and connectivity among the processing elements. In these networks, information is represented using populations of different types of interacting agents, such as molecules, cells, \citep{Turing1952-bj} or even animals \citep{Lotka1926-ei}, and the structure of the computational process is encoded in the interaction-reaction rules between these agents. Chemical Reaction Networks (CRNs) are a notable example of computational systems of this type, capable of making complex decisions and adapting even under the assumption that individual computing elements undergo completely chaotic Brownian motion. In this work we aim to use differentiable optimization to automatically design task-specific networks of this type.

Assuming mass-action kinetics and that individual chambers are well-mixed, \cite{CME} naturally defines a differential equation for modelling the dynamics of a given CRN. The modelled variables are concentrations of participating chemical components, and their rates of change are defined by the structure of the reaction network and current concentrations of the reactants, catalysts, and inhibitors.

The computational power of such networks has been proven to be Turing Complete \citep{erikBoolean} and is thus sufficient for representing arbitrary computation, such as, for instance, computations representable by a Boolean formula. 

While the possibility of computation with CRNs has long been enticing, implementations have generally involved significant complexity \citep{directCRN}. In recent years, DNA strand displacement has been demonstrated as a viable implementation mechanism for CRNs \citep{erikDNAforCRN}.



\subsection{Computing with Chemical Reaction Networks}

In this work we focus on systems that have transition rules of the following form: $A+C \xrightarrow{k} B+C$, meaning that substance $A$ gets transformed into $B$ after interaction with $C$, which acts as a catalyst. The rate at which the chemical reaction occurs can be expressed as $kAC$, where $A$ and $C$ are the current concentrations of the reactant and the catalyst and $k$ is a reaction coefficient. We are going to use a more compact notation for this type of reaction: $A \xrightarrow{k,C} B$. This choice of elementary reaction type gives us an ``agent-centric'' view of the system where each agent makes independent decisions about its state after an interaction with another agent.

\begin{figure}
     \centering
     \begin{subfigure}[b]{0.15\textwidth}
         \centering
         \begin{align*} 
            A \xrightarrow{k_1, B} B \\
            B \xrightarrow{k_2, C} C \\
            C \xrightarrow{k_3, A} A
         \end{align*} 
         \caption{reaction rules}
         \label{fig:ocs-react}
     \end{subfigure}
     \begin{subfigure}[b]{0.3\textwidth}
         \centering
         \begin{spreadlines}{0.8em}
         \begin{align*} 
            A' = k_3 C A - k_1 A B \\
            B' = k_1 A B - k_2 B C \\
            C' = k_2 B C - k_3 C A
         \end{align*} 
         \end{spreadlines}
         \caption{corresponding ODE system}
     \end{subfigure}
     \begin{subfigure}[b]{0.2\textwidth}
         \centering
         \includegraphics[width=\textwidth,trim={1.3cm 1.3cm 1.3cm 1.3cm}]{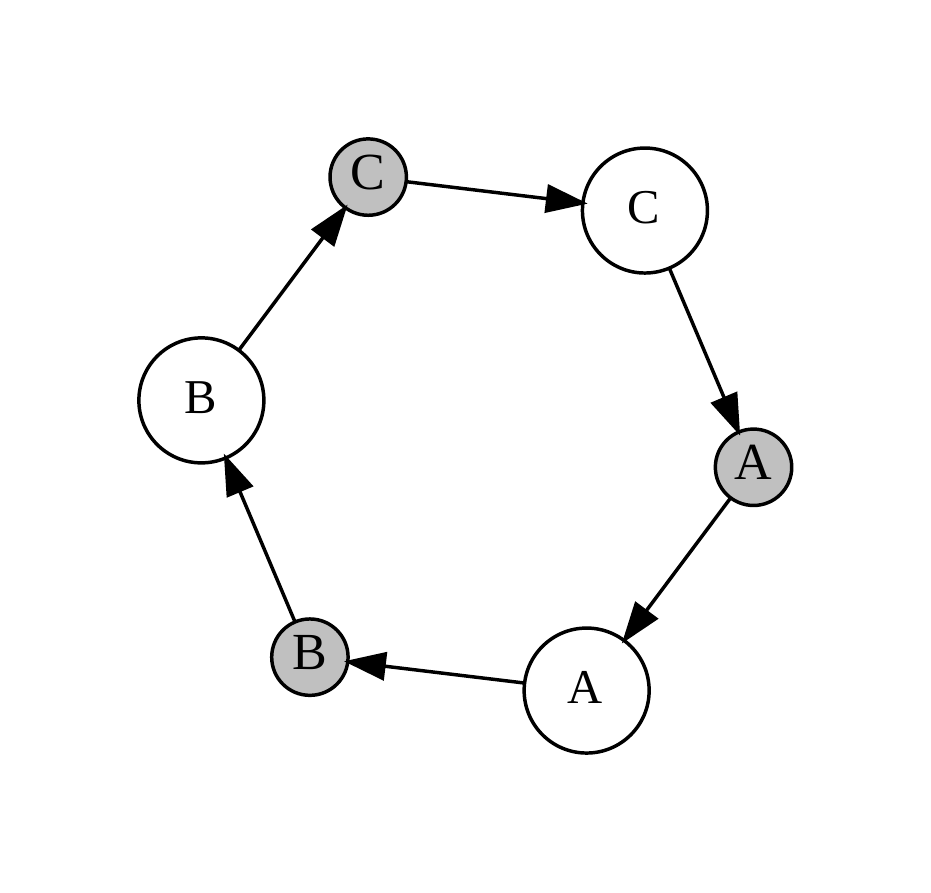}
         \caption{reaction graph}
         \label{fig:ocs-graph}
     \end{subfigure}
     \begin{subfigure}[b]{0.3\textwidth}
         \centering
         \includegraphics[width=\textwidth]{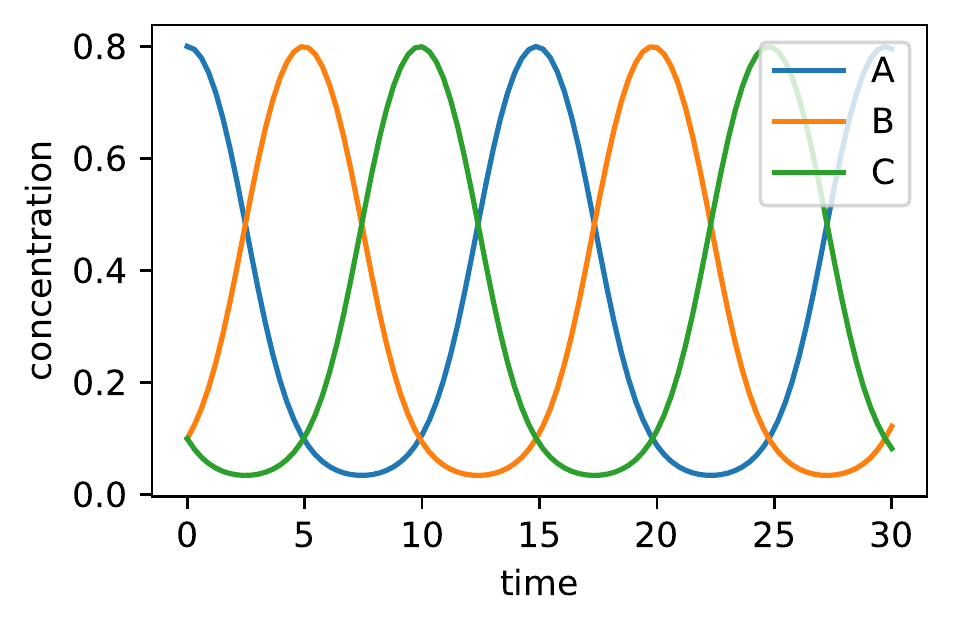}
         \caption{system dynamics}
         \label{fig:ocs-plot}
     \end{subfigure}
\caption{Reaction Network example: three-phase oscillator. Graph (\subref{fig:ocs-graph}) shows reactants as white nodes and reactions as grey nodes. Grey node labels denote the catalyst, that activates a particular reaction. Plot (\subref{fig:ocs-plot}) shows the evolution of the oscillator system in case $A_0=0.8$, $B_0=C_0=0.1$, $k_{\{0,1,2\}}=1$}
    \label{fig:osc}
\end{figure}

Figure \ref{fig:osc} shows a simple CRN oscillator composed of such reactions. It is easy to transform reaction rules into an ODE-system, where each reaction decreases the concentration of its input and increases the concentration of its output. We can think of concentrations as variables, where reactions are ``gates'' that continuously modify the variable values. The language of chemical reactions is surprisingly versatile and powerful. For example, it is possible to simulate the function of an arbitrary Boolean circuit using a network constructed from reactions of the type described here \citep{erikBoolean}. Various analog circuits, such as oscillators, approximate majority computators, or even 3D renderers \citep{3DCRN}, can be constructed as well.

Given the rise in popularity of artificial neural networks (ANN), it is not surprising that chemical networks were adopted to perform ANN inference. Typically, CRNs are constructed to compute results of formulas composed of traditional NN building blocks, such as matrix multiplications and element-wise non-linearities. In contrast, in this work we explore the possibility of direct application of backpropagation gradient-based optimization to find the CRN network structure and parameters for solving a particular problem. 

Numerical optimization has already been applied to determining the parameters of physical systems that satisfy real world measurements from various processes \citep{sindy} (model identification). In this work we focus on synthesising \textit{new models} given the specification expressed with an objective function. From that perspective, contributions of this paper can be summarized as follows:

\begin{itemize}
    \item Define an efficient parameterization and a training procedure that enables differentiable optimization of CRNs having a particular structure.
    \item Demonstrate that this procedure can be used to synthesize compact reaction networks that perform computational tasks specified by a provided objective function.
\end{itemize}

\section{Differentiable Reaction Networks}

In this section, we describe a possible differentiable representation of a CRN system. Consider $N$ components that undergo reactions of the form $X \xrightarrow{k,Z} Y$, where $X$, $Y$ and $Z$ are arbitrary (may even be repeating) components from these N. The total number of possible reactions (including the trivial $X \xrightarrow{Z} X$) is therefore $N^3$. We may represent any such $N$-element reaction network using a 3-dimensional tensor $T$ of shape $N \times N \times N$, where dimensions correspond to the reaction catalyst, input and output respectively. We are going to call this the \textit{reaction tensor}. For example, the following tensor represents the oscillator system shown on the Figure \ref{fig:osc}:

\begin{equation*}
\begin{tabular}{ c c c }
 $T_{0,*,*}$ & $T_{1,*,*}$ & $T_{2,*,*}$ \\ 
$\begin{bmatrix}
0 & 0 & 0\\
0 & 0 & 0\\
k_3 & 0 & -k_3
\end{bmatrix}$
& 
$\begin{bmatrix}
-k_1 & k_1 & 0\\
0 & 0 & 0\\
0 & 0 & 0
\end{bmatrix}$
& 
$\begin{bmatrix}
0 & 0 & 0\\
0 & -k_2 & k_2\\
0 & 0 & 0
\end{bmatrix}$
\end{tabular}
\label{eq:T}
\end{equation*}

We refer to elements of this tensor as $T_{c,a,b}$, where each slice $c$ corresponds to one catalyst, and each row $a$ to one reaction input. Elements of each row express the change of the concentration of each substance caused by the reaction of a particular input-catalyst pair. For example, the row $T_{1,0} = [-k_1,  k_1,  0]$ means that when the 1\ts{st} component (B) catalyses the 0\ts{th} component (A), A gets removed and B gets added with the rate $k_1$, which corresponds to the reaction $A \xrightarrow{k_1, B} B$ from the Figure \ref{fig:ocs-react}. Rows of $T$ have a meaning, similar to the rows of a Stoichiometric matrix, but also encode reaction rates along with reactants and products. Consider a vector $\mathbf{x} = [x_0, ..., x_{N-1}], \forall i : x_i \ge 0$ that encodes current concentrations of $N$ chemicals. We can now express the rate of change of its components over time as easily as 

\begin{equation}
x'_b = \sum_{a, c} x_a x_c T_{c,a,b}
\label{eq:react_ode}
\end{equation}

We need to impose some constraints on the tensor $T$ on make sure that it represents the right type of reactions:

\begin{itemize}
    \item $\forall a, c :  \sum_b T_{c,a,b} = 0$ \; --- all rows sum to zero to conserve the mass;
    \item $\forall a, c : T_{c, a, a} \le 0$ \; --- only reactant may get consumed by the reaction;
    \item $\forall a, b, c : a \ne b \implies T_{c, a, b} \ge 0$ \; --- reaction outputs must be non-negative;
\end{itemize}

Note that this formulation allows multiple possible outputs for the same input. For example, let's suppose that $T_{1, 0} = [-1, 0.8, 0.2]$. This may be interpreted as two reactions $A \xrightarrow{0.8, B} B$ and $A \xrightarrow{0.2, B} C$ running in parallel. We can also think of molecules as \textit{stochastic finite state machines} (FSM), that make a random decision on which state to take upon interaction with another molecule. This view inspired us to use the following differentiable representation of the reaction network, which maintains the properties listed above. We construct $T$ from a logit parameters tensor $W \in \mathbb{R}^{N \times N \times N}$ this way:

\begin{equation}
\begin{split}
    P_{c,a,b} &= \underset{b}{\mathrm{softmax}}(W_{c,a,b}) \\ 
    T_{c,a,b} &= P_{c,a,b}-I_{a,b}
\end{split}
\end{equation}

Rows of $P$ are probability distributions over the resulting molecule states for each possible input-catalyst pair. Rows sum up to one, so we can subtract the identity matrix $I$ that spans axes $a$ and $b$ to obtain the reaction tensor $T$.

Once we defined the differentiable representation (\ref{eq:T}) for the coefficients of the reaction network ODE system (\ref{eq:react_ode}) and selected the initial conditions $\mathbf{x}(0)$, we may plug the equation into a differentiable ODE solver. Gradient-backpropagation through the ODE solver may be either performed directly, or by using the adjoint state method method \citep{neuralODE}.

In following sections we are going to explore a number of different optimization objectives that are expressed in terms of the behavior of the ODE system defined by the reaction network.

\subsection{Sparsity-inducing regularization}

It's often desirable to keep the resulting network as simple as possible. One possible definition of simplicity is the number of different reactions that are possible within a network. Our key motivations for reducing the network complexity are feasibility of physical implementation and interpretability of the network structure.

We consider each unique combination of catalyst, reactant and product as one reaction. The maximum number of possible reactions our model allows for a system of $N$ species is $N^3-N^2$, where $N^2$ accounts for the excluded "no-op" $X \xrightarrow{Y} X$ reactions. In our experiments we use a few of strategies to reduce the number of reactions in the CRN. We pick a threshold value $k_\text{min}$ and ignore all reactions that have a smaller rate. We can do so by setting the corresponding elements of tensor $T$ to zeros. Note that we also have to adjust negative diagonal ($T_{c,a,a}$) elements to make sure that each row still has a zero sum. We call this sparsified tensor as $T^{k_\text{min}}$.

\paragraph{Regularization losses} 
We use a number of additional training loss terms to steer the optimization into finding the reaction tensors $T$ that have a larger number of near-zeros values that can be discarded:

\begin{center}
\begin{tabular}{ c c c } 
 $\displaystyle L_{L1}=\frac{1}{N^3}\sum_{c,a,b} |T_{c,a,b}|$ & 
 $\displaystyle L_H=-\frac{1}{N^3}\sum_{c,a,b} P_{c,a,b} \mathrm{log}(P_{c,a,b})$ & 
 $\displaystyle L_I=-\frac{1}{N^3}\sum_{c,a,b} P_{c,a,b}^2$
\end{tabular}
\end{center}

$L1$-regularization ($L_{L1}$) is a common ways to promote sparsity of optimized parameters. The corresponding loss term boils down to computing the average of absolute values of elements of $T$.

Another approach to regularization is inspired by the stochastic state machines interpretation of molecules. We would like to reduce the number of stochastic reactions, i.e. reactions which produce more then one possible output for a given reactant-catalyst pair. One way of steering optimization towards such networks is decreasing the entropy ($L_H$) of rows of the tensor $P$, which can be interpreted as a stochastic transition table of FSMs that represent molecules. A similar effect can also be achieved by maximizing the so called \textit{informational energy} ($L_I$) \citep{informational_energy}.

The set of used loss terms and their weights vary from experiment to experiment and described in corresponding appendix sections.

\paragraph{Sparsified training} In some experiments we observed that post-training removal of small rate reactions from the network may lead to substantial difference with the behaviour seen during training. For example, in dynamics matching experiments (section \ref{sec:dyn_match}), the frequency of sparsified learned oscillators diverged from the objective. This lead us to the idea of accounting for sparsification during the network training. We experimented with two strategies of training sparse networks. The first strategy is to use $T^{k_\text{min}}$ instead of $T$ on forward training pass, but propagate gradients back to $T$ as if small value masking didn't happen\footnote{This is often achieved with the \texttt{stop\_gradient} trick: $T_\text{train} = T + \mathrm{stopgrad}(T^{k_\text{min}}-T)$}. An alternative approach involves computing the target specific objective function twice, using the original and sparsified networks, and adding the results together: $L = \mathrm{Loss}(T) + Loss(T^{k_\text{min}})$

\section{Waveform matching}
\label{sec:dyn_match}

\begin{figure}
    \centering
    \includegraphics[width=\textwidth]{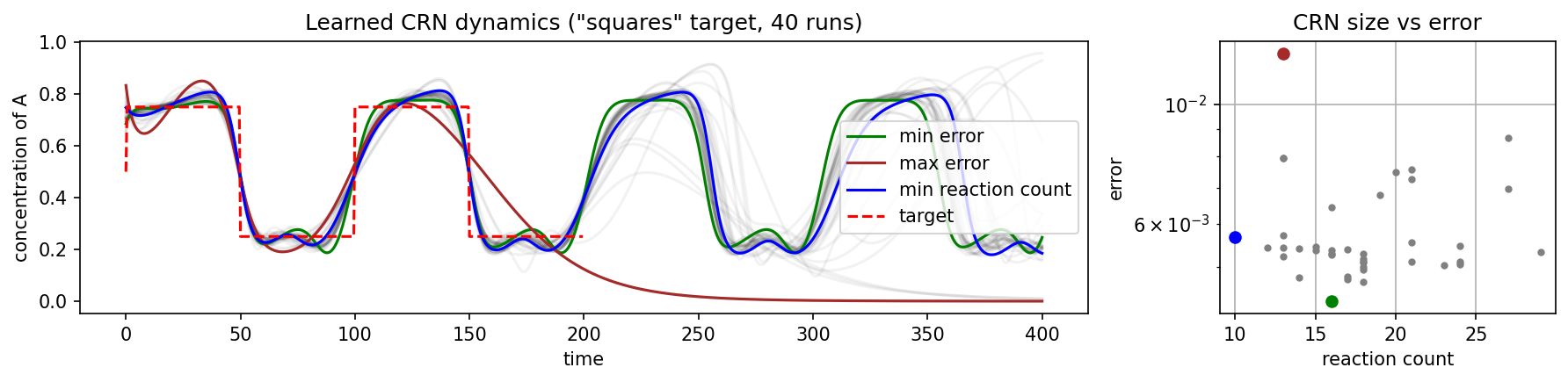}
    \includegraphics[width=\textwidth]{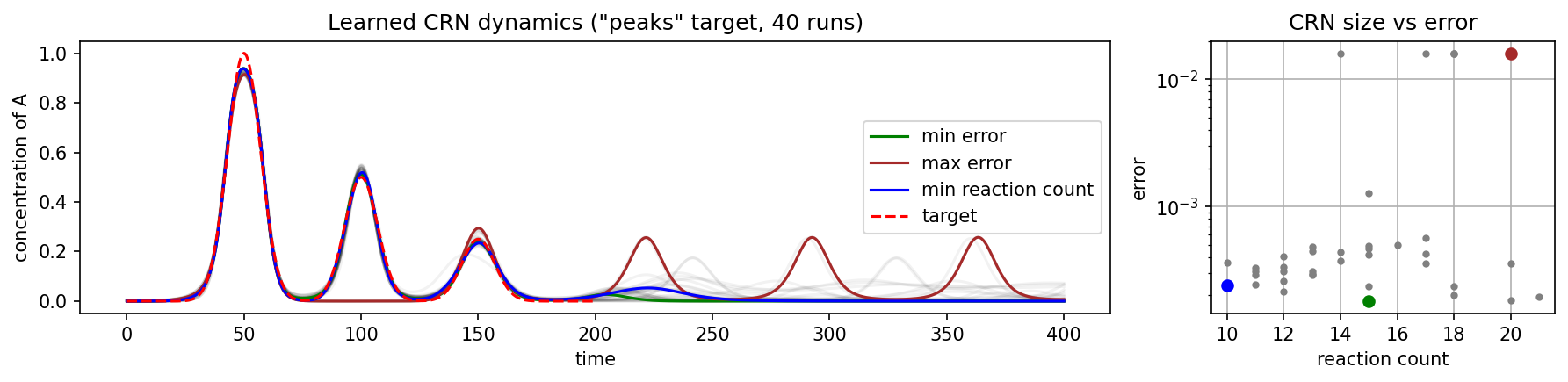}
    \caption{Results of training 5-component CRNs to reproduce two different temporal target patterns. We trained 40 CRNs for each pattern using different random initialization. Most runs converged to solutions that produced reasonable approximations of the target waveform. We observed large variance in numbers of reactions constituting sparsified CRNs ($k_\text{min}=10^{-3}$). A large fraction of "squares" target runs converged to the oscillating solution, although it was not explicitly required by the training objective.}
    \label{fig:dyn_matching}
\end{figure}

In this section we explore the capability of differentiable optimization to find reaction networks, which have specific temporal dynamics. Consider a scalar function $f(t)$ defined on the range $[0, t_{\text{max}}]$. We would like to find a reaction network $T$ and initial conditions $\mathbf{x}(0)$, so that the temporal dynamics of concentration of one of the chemical components (e.g. $x_0$) matches the target function $f$ as closely as possible. We define the objective in the following way:

$$ L_f = \frac{1}{t_\text{max}}\int_0^{t_\text{max}} (x_0(t)-f(t))^2 dt
\; \approx \; \frac{1}{n_t}\sum_{i=0}^{n_t} (x_0(t_i)-f(t_i))^2
$$
where $t_i$ values are evenly spaced over the $[0,t_{max}]$ interval.

We study two different examples of target function dynamics: square wave and three peaks of decreasing intensity. Figure \ref{fig:dyn_matching} shows target waveforms along with the behaviours of 40 independently trained networks that were using different random parameter initialization. The target loss is applied over the interval $[0,t_{max}]$. We evaluated resulting CRNs on a twice longer time interval to see how the learned behaviour generalizes outside of training time frame. We observed that the proposed procedure is capable to discover compact CRNs that demonstrate an approximation of the target dynamics in concentrations of one of the chemical components.

\section{Functional networks}
In this section we explore the capacity of learned CRNs to find approximations to some simple functions. We investigate approximations to binary functions $f: \mathbb{Z}_2^i \rightarrow \mathbb{Z}_2^j$ as well as functions of the form $f : \mathbb{R}_{>0}^i \rightarrow \mathbb{R}_{>0}^j$. We refer readers to the supplementary materials further examples of learned functions, such as Analogue-to-Digital converters. Given both  measurable and controllable quantities in CRN are real-valued, non-zero concentrations of a chemical, and we use a similar approach to \cite{oxford} to map the space of concentrations of indicator chemicals to high and low signals in a binary setting.

\subsection{Logic Gates}

There has long been an interest in implementing Boolean operators as reaction networks. Several successful hand-engineered implementations have been demonstrated \citep{erikBoolean,oxford}, with varying properties and encoding schemes. The CRN design in \cite{oxford} additionally has the key property of \textbf{reusability} - the control chemicals can be changed and the CRN responds appropriately, updating its output. This property is non-trivial as it requires a network to be able to maintain a state, as opposed to use-once circuits.

We take inspiration from \cite{oxford} and demonstrate that our proposed method can learn CRNs that approximate Boolean functions, and can learn them in a \textbf{reusable} fashion. We use the same number of CRN chemicals per Boolean operator, and the same encoding scheme for inputs and outputs. Additionally, we initialize non-indicator and non-input chemicals to the same concentrations as in \cite{oxford}. We note that the values of these auxiliary chemicals are not readily included in \cite{oxford}, but we infer them to the best of our ability from the time-concentration graphs.  

\subsubsection{Dual Rail Encoding}

We use dual rail encoding as in \cite{oxford}. Each input and output variable $X$ is represented by two unique complementary chemicals, $X_{hi}$ and $X_{lo}$, whose concentrations signal the state of variable $X$. We consider $X = 1$ i.f.f. $X_{hi} >= 1.0 - \epsilon$ and $0 <= X_{lo} <= \epsilon$, and $X=0$ otherwise. We choose $\epsilon = 0.1$, however in practice when designing loss functions, we encourage $X_{hi}$ and $X_{lo}$ to be as close as possible to one of the two desired states $(1, 0)$ or $(0, 1)$ at measurement time and often find that learned solutions converge to states much closer than $\epsilon$. 

\subsubsection{Target \& Training}

We allocate the $N$ chemicals used in the CRN into the inputs, $\textrm{IN} := \{X_{hi}, X_{lo}, Y_{hi}, Y_{lo}\}$, outputs $\textrm{OUT} := \{Z_{hi}, Z_{lo}\}$ and $N-6$ auxiliary chemicals $\textrm{AUX} := \{A, B, C, D ...\}$. We explicitly prevent backpropagation into indices of our reaction tensor $T$ which would consume or produce the any chemical in $\textrm{IN}$, ensuring that these act as fixed control chemicals, and can only influence the CRN dynamics as catalysts. 

We independently train three CRNs to learn the Boolean operators "\textbf{AND}", "\textbf{OR}" and "\textbf{XOR}". In each case, we use as many $\textrm{AUX}$ chemicals as used in \cite{oxford}, which is 3, 3 and 4, respectively. We train our CRN largely using the method outlined in \ref{sec:dyn_match}, with a few caveats. For each operator, we generate a training set of initial concentrations and timed transitions of the input chemicals $\textrm{IN}$ and matching desired outputs $\textrm{OUT}$. Inputs $(X, Y)$ can be one of $\{(1, 0), (0, 1), (1, 1), (0, 0)\}$, so the set of transitions consists of $2^4$ possible transitions (e.g. $(1, 0) \rightarrow (1, 1)$). During training, we run the CRN for $T=800$ time, and introduce a transition in the input after every $T//4$, i.e. at $T = 200, 400, 600$. Each batch entry in our training set covers four transitions. 

We then impose the aforementioned waveform loss, only on the output chemicals, but over a period of $T//8$ prior to the next transition, encouraging the CRN to converge to the correct output for the given inputs just before the next transition. We use an L1 instead of L2 loss to further encourage stability in the outputs.  

\subsubsection{Results \& Verification}

\begin{figure}[h]
 \includegraphics[width=0.33\textwidth]{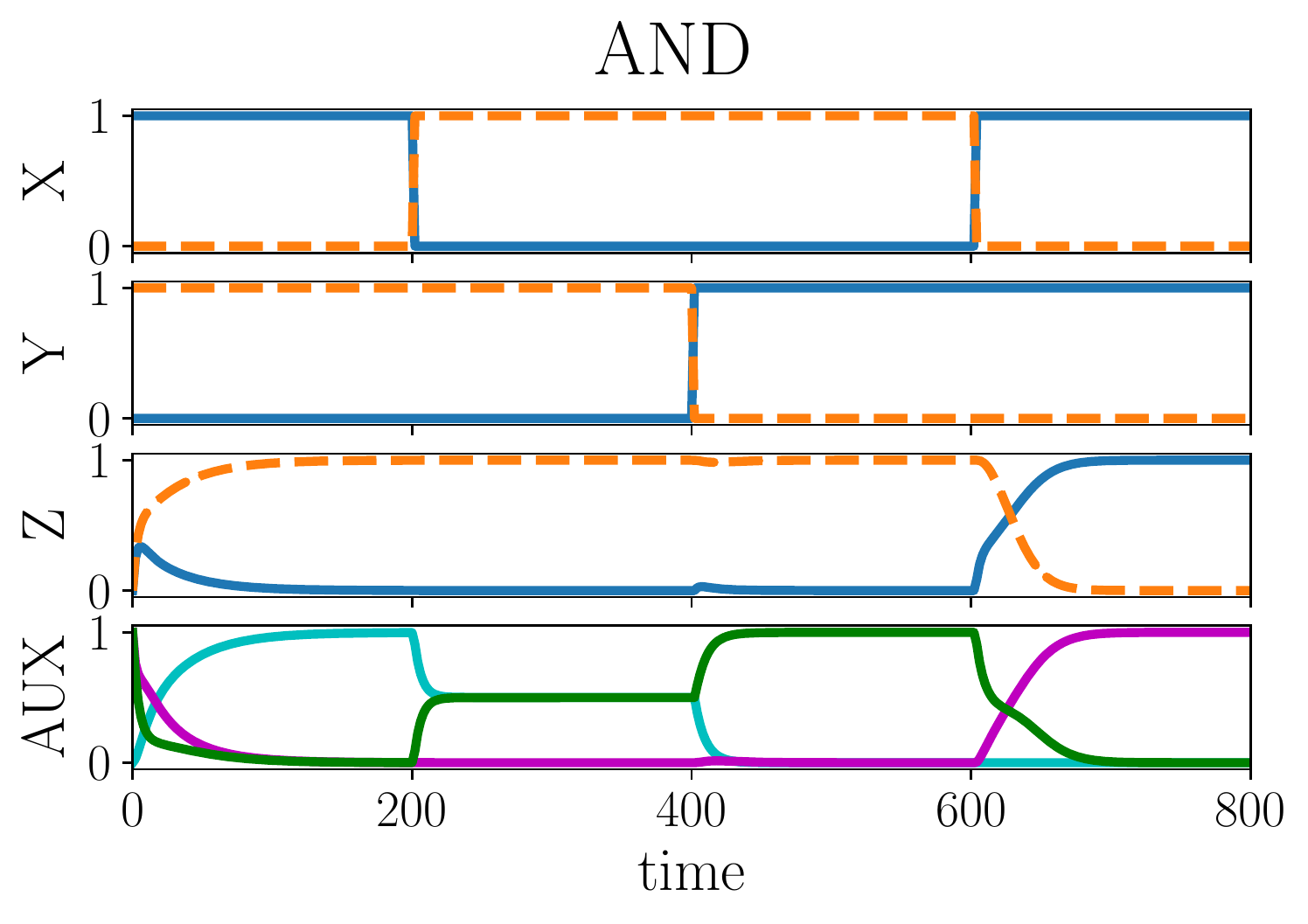}
 \includegraphics[width=0.33\textwidth]{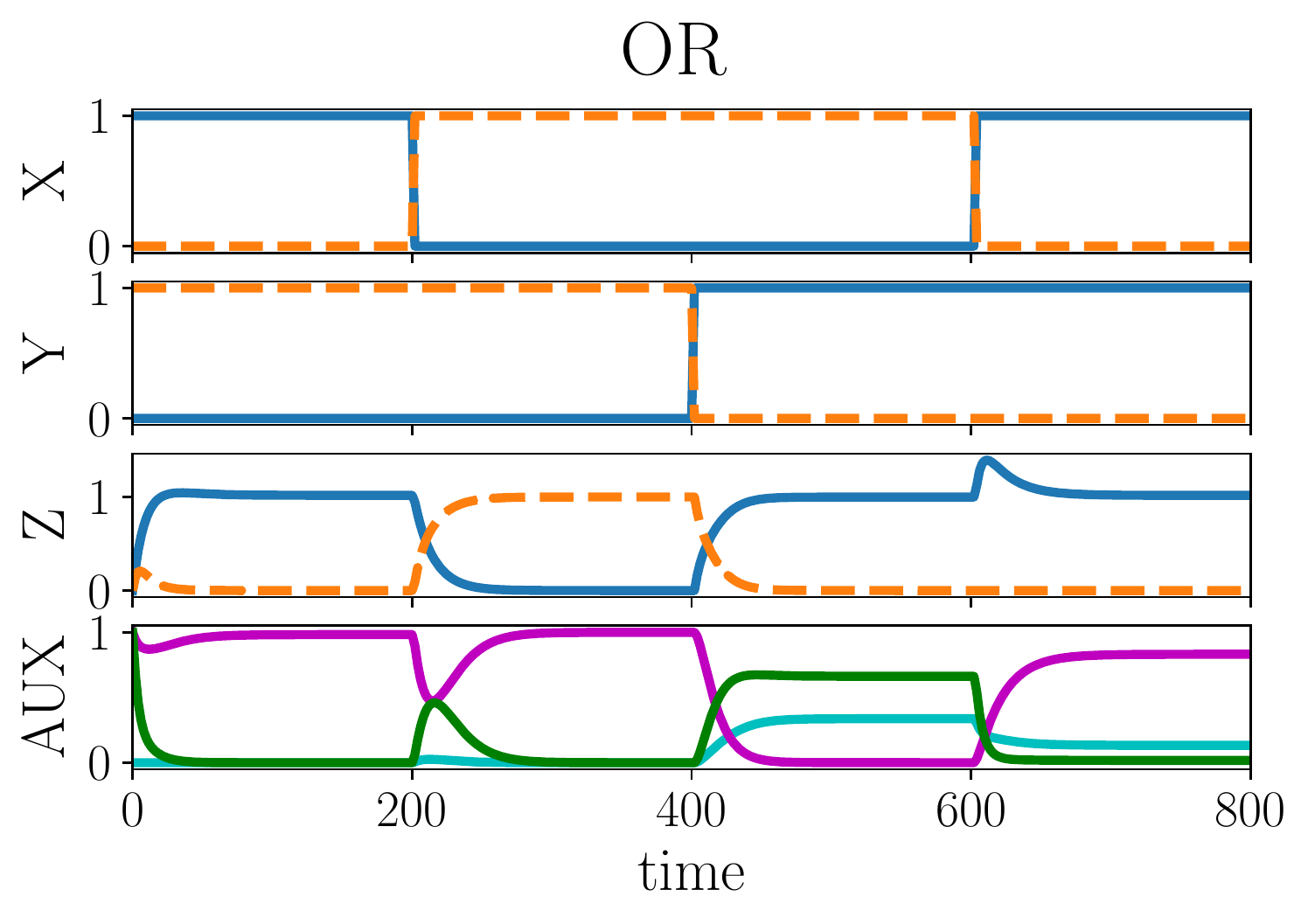}
 \includegraphics[width=0.33\textwidth]{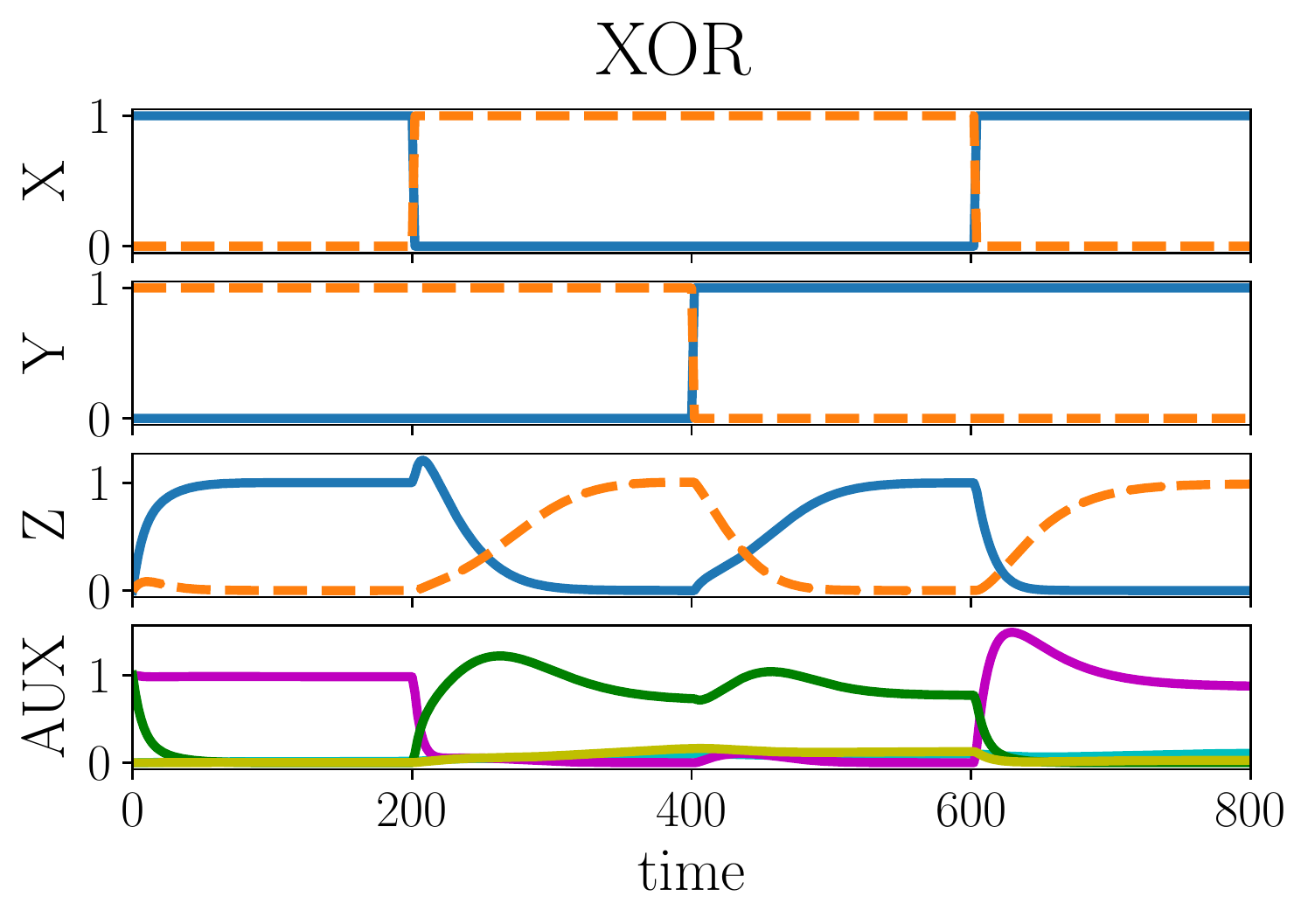}
 \caption{Dynamics of learned CRNs approximating the Boolean operators AND, OR and XOR.} 
    \label{fig:OR}
\end{figure}

We refer to figure \ref{fig:OR} for a sample of the dynamics of the learned CRNs over time. AND, OR and XOR CRNs consist of 20, 16 and 14 reactions with a rate $> 0.1$, respectively, which compares favourably with the functionally equivalent hand-designed CRNs in \cite{oxford}, with 7, 7 and 12 reactions, respectively.

In \cite{oxford}, correctness of the designed CRN is proved through a combination of informal reasoning about the circuit, simulation of the circuit using Visual GEC \citep{visualGEC} under both deterministic and under stochastic conditions, as well as formally verifying the circuit using PRISM \citep{PRISM}. 

 The mechanics of our learned circuits are non-trivial to reverse-engineer and formally verify in similar fashion. Instead, we perform two tests on the stability of the learned CRN. Firstly, we evaluate the behaviour of the CRN on the a modified training dataset iterated for $T_{eval} = 100*T_{train}$, with the transition points placed accordingly every $T_{eval}//4$. Secondly, we iterate the CRN again for $T_{eval}$ steps, but continuously uniformly sample the "time to next transition" from ${U_{[T_{train}//4,T_{train}//2]}}$, as well as randomly sample the next input state.
 
All our learned Boolean CRNs output the correct values (under our $\epsilon$ definition) during these tests, suggesting convergence to a stable point.

\subsection{Seven segment display mapping}

\begin{figure}
 \centering
 \includegraphics[width=0.95\textwidth]{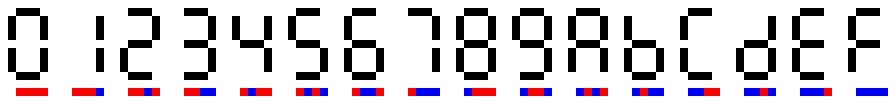}
 \caption{Seven-segment digit mapping task. The 4 input bits (red and blue squares) are mapped to 7 output gates each representing a segment (black line) in a hexadecimal display.}
    \label{fig:ssd_task}
\end{figure}

One more complex logical mapping is the Seven-segment digit mapping (Figure~\ref{fig:ssd_task}), where the combinations of 4 input bits are mapped to 7 output segments activations. We define a \textit{low} and \textit{high} floating point parameters, representing the initial values of 4 input chemicals for the 0 and 1 input case respectively. For instance, the input encoding "1010" is mapped to four chemicals initialized as follows: \textit{(high,~low,~high,~low)}. We decided to keep the total mass of chemicals equal across different inputs. To do that, we add another input whose initial value is equal to $1$ minus the sum of the 4 input chemical concentrations for that instance. Therefore, the resulting input encoding consists of $n+1=5$ chemicals. The output is defined by the final value of 7 output chemical distributions (note we \textit{do not} request any target value for the final input chemicals). We choose to train the task with a \textit{squared hinge loss} on the output chemicals:

\begin{equation}
\begin{matrix}
\textit{TranslateAndScale}(x) = (2x - \textit{low} - \textit{high})/(\textit{high} - \textit{low}) \\
\textit{SquaredHingeLoss}(x_i, y_i) = (\text{Max}(0, 1 - \textit{TranslateAndScale}(x_i)\cdot y_i))^2
\end{matrix}
\label{eq:sqhinge}
\end{equation}

where the vector $y$ has its value set to $-1$ and $+1$ for output values of $0$ and $1$ respectively. The squared Hinge loss effectively penalizes the output chemicals if they do not get lower than \textit{low} or higher than \textit{high} if the target output is 0 or 1 respectively. We choose to apply this loss on the latter half of the time unfolding (as opposed to only the final result), encouraging a more stable final configuration. Finally, we add 4 more chemicals (initialized to zero) as auxiliary channels. To evaluate whether the task is solved, we then threshold the output concentrations and consider them 0 or 1 if they are below or above this threshold. The arbitrary choice of a midpoint threshold = $(\textit{high} + \textit{low})/2$ appears to work well and with it we achieve a perfect fit on this task.

\subsection{Single-chamber winner-takes-all}\label{sec:scwta} 

We qualify this task with "single-chamber" as we present a more complex case in the next section. The premise of the task, also sometimes referred to "approximate-majority", is to treat two chemicals ${A, B}$ as both input and output, with initial concentrations $A_0, B_0 \in [0, 1]$, and to have the desired final, converged, state of our chemicals to be:

$$
\lim_{t \to \inf} (A_t, B_t) = 
\begin{cases}
(1.0, 0.0) & A_0 > B_0\\
(0.0, 1.0) & A_0 < B_0
\end{cases}
$$

\begin{figure}[h]
 \raisebox{3.0cm -\heightof{\textbf{(a)}}}{\textbf{(a)}} \includegraphics[trim={0.0cm 0.0cm 0.0cm 0.0cm},clip,width=0.95\textwidth]{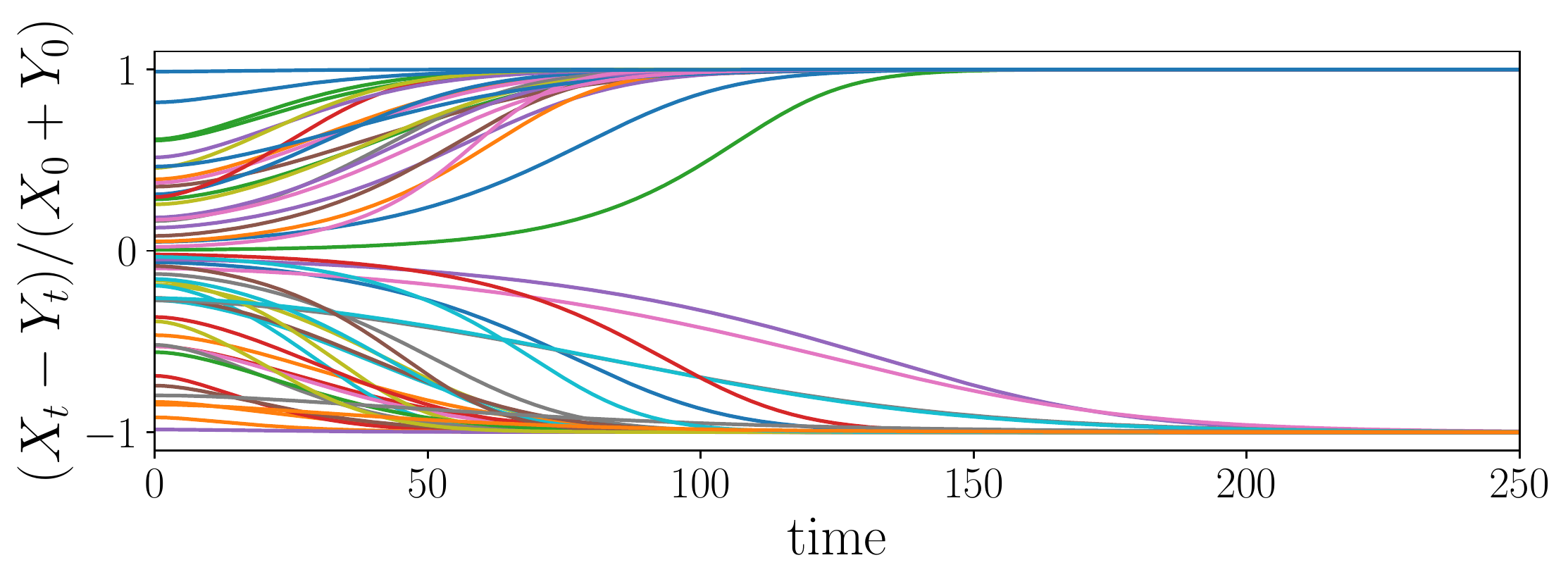}\hspace{-6.80cm} \raisebox{1.3\height}{\raisebox{0.95cm -\heightof{\textbf{(b)}}}{\textbf{(b)}}\includegraphics[width=0.38\textwidth]{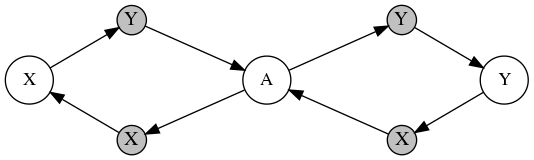}}
 

 \caption{\textbf{(a)} Learned single-chamber winner-takes-all CRN. Each line corresponds to the evolution of one instantiation of the chamber. Note the magnitude of the measured quantity does not converge exactly to 1.0, due to the small-magnitude secondary reactions. When initial concentrations are very close (i.e. the measured quantity is close to $0.0$) the secondary reactions cause incorrect convergence. \textbf{(b)} Graphical representation of the learned reaction network, showing only reactions rates $k_r > 0.05$ (n.b. depicted reactions all have $k_r = 1.0$). The sparsified learned reaction network exactly matches the approximate majority network derived in \cite{AMpaper}} \label{fig:wtaplot}
\end{figure}

This definition also implies convexity in the dynamics of the CRN, but we don't explicitly enforce this. Using \cite{oxford} as a heuristic for the upper bound on the number of chemicals required for such a reaction-network, we design the reaction with only one additional auxiliary chemical, initialized with a concentration of 0. We train the model by initializing it with sampled $A_0, B_0 ~ U_{[0,1]}$, and apply an L2 loss at time $T=200$, penalising the deviation from the desired states defined above. In figure \ref{fig:wtaplot} we show the dynamics of the reaction network, as well as it's time evolution for various initialisations of $X_0$ and $Y_0$, plotting the evolution of the difference between the two concentrations normalized by the magnitude of their sum at $T=0$. We also graphically visualise the sparsified CRN.

We are keen to note that the sparsified version of the learned reaction network for approximate majority is \textbf{identical} to the hand-designed, formally verified one in \cite{AMpaper}.

\section{Multi-chamber reaction-diffusion models}

We now introduce the concept of \textit{membranes} and their related diffusion of certain chemicals. Traditionally, CRNs have often had their reaction component paired with a \textit{diffusion} component, where chemicals on a space or through a membrane would naturally diffuse with varying rates \citep{Turing1952-bj, Kondo2010-kb, Mordvintsev2021-eo}. In this paper, we focus on diffusion occurring on permeable membranes, and more specifically on \textit{rings} of chambers, where every chamber has a right and a left neighbour (with the exception of the case with only two chambers where there is only one neighbour). We extend our ODE system to take into account the contribution on the change of rate for any chemicals passing through membranes. We construct a diffusion trainable vector $V \in \mathbb{R}^N$ (we only need one single value for each chemical to represent their diffusion through a membrane) and construct the diffusion rate vector: $D = \mathrm{Sigmoid}(V)$. Now, we can create systems with different chambers, separated by membranes. Each chamber has their own concentration of chemicals that react among themselves. Different chambers connected by a membrane diffuse chemicals at different rates, based on the vector $D$. The rate of change of each chemical component now becomes:

\begin{equation}
{x^i_b}' = \sum_{j}^{M_i} (x^j_b - x^i_b) D_b + \sum_{a, c} x_a x_c T_{c,a,b} 
\label{eq:main_ode}
\end{equation}

where $i$ is a chamber identifier and $M_i$ is the set of chambers connected to i. For this task every membrane is identical, but this system can generalize to different diffusion rates for different membranes if needed.

\subsection{Multi-chamber winner takes all}

We revisit the winner takes all task we introduced in section~\ref{sec:scwta} and render it a multi-chamber task. In this version, we randomly initialize a chemical A within two values \textit{low} and \textit{high} across different chambers. The task is to suppress A on all the chambers where their concentration was not the highest, and highlight A on the winner chamber. This task can also be seen as a variant of \textit{leader election} in anonymous rings \citep{Xu2015-zi}, where the leader needs to suppress all other nodes, with the added complexity that a leader must be chosen based on the input configuration of a specific group of chemicals. The capacity of performing leader election is an extremely important feature of most biological systems, as it enables differentiation of roles. For instance, analyses of \textit{Drosophila} have been demonstrated to perform leader election routines \citep{Afek2011-lf, Barad2011-fc, Barad2010-gl, Jacobsen1998-sl}.

\paragraph{Task description}

We construct $n$ chambers connected as a ring through membranes sharing the same diffusion vector $D$. During training we set $n=5$ exclusively and we evaluate for more out-of-training configurations. The chemical A is randomly initialized within $\textit{low}=0.01$ and $\textit{high}=0.9$ in each chamber. We also initialize three more auxiliary chemicals (B,C,D) to zero everywhere and enforce the input configuration to have a total concentration (sum of all chemicals) for each chamber to be equal to 1. We do so by adding a chemical E initialized as $E^i = 1. - A^i$ for each chamber i. We observed this initialization to be critical for a successful training of this system. We apply a squared Hinge loss (Equation~\ref{eq:sqhinge}), with lower and upper bound of \textit{low} and \textit{high} respectively, for all steps after the first $1/3$rd, encouraging the model to find a more stable final configuration.

\paragraph{Results}

\begin{figure}
     \centering
     \begin{subfigure}[t]{0.35\textwidth}
         \centering
         \includegraphics[width=\textwidth]{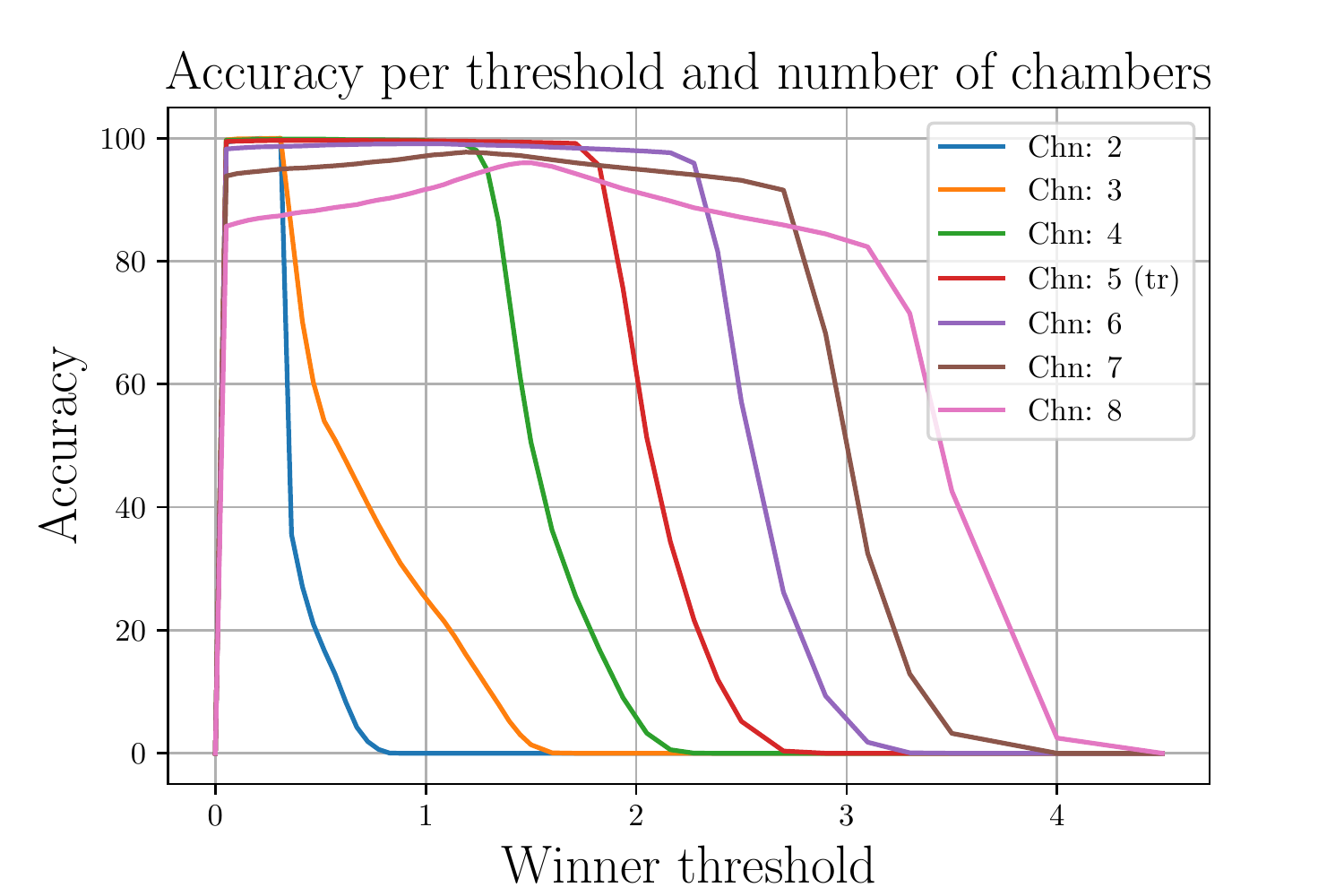}
         \caption{Accuracies per winner threshold}
         \label{fig:accuracy_per_thr}
     \end{subfigure}
     \begin{subfigure}[t]{0.60\textwidth}
         \centering
         \includegraphics[width=\textwidth]{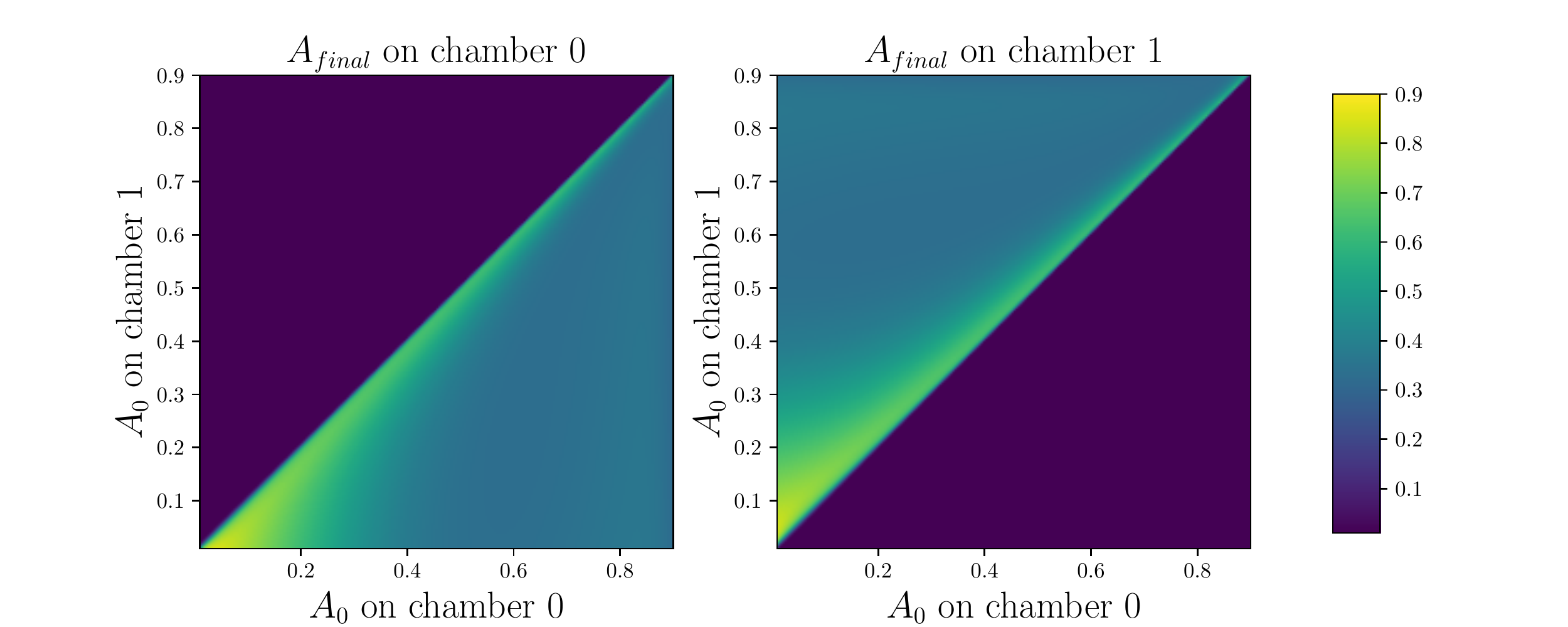}
         \caption{Input/Output distributions with two chambers}
         \label{fig:binary_leader_d}
     \end{subfigure}
\caption{Plot (\subref{fig:accuracy_per_thr}) shows accuracy results on an eval dataset (n=10000) for varying values of winner thresholds and number of chambers (Chn). The number of chambers used during training is 5. Plot (\subref{fig:binary_leader_d}) shows the complete input-output mapping of the resulting network for the case with two chambers.}
    \label{fig:accuracy-plot}
\end{figure}

Given a \textit{winner threshold t}, a batch of final configurations x of chemicals A per chamber and a batch of target winner and losers configurations y, we define $\textit{accuracy}(t, x, y)$ as the percentage of instances $b$ where only the winner chemical on $y_b$ is above the threshold t in $x_b$.
\begin{table}
\caption{Accuracies on the multi-chamber winner takes all task.}
\label{tab:wta}
\centering
\begin{tabular}{*{9}{l}}  
\toprule
&\multicolumn{8}{c}{Number of chambers}\\
                  & 2    & 3    & 4    & 5    & 6    & 7    & 8    & all \\
\midrule
Threshold         & 0.21 & 0.31 & 0.36 & 0.52 & 1.04 & 1.19 & 1.5  & 0.31\\ 
Eval Accuracy (\%)& 99.95& 99.99& 99.87& 99.63& 99.10& 97.73& 96.01& 97.19  \\ 
Test Accuracy (\%)& 99.97&100.00& 99.88& 99.68& 98.87& 97.63& 95.36& 97.27  \\ 
\bottomrule
\end{tabular}
\end{table}

Figure~\ref{fig:accuracy_per_thr} shows the different accuracy results for varying t and different numbers of chambers. Losers are consistently suppressed to $\sim0$ for all chambers, while more chambers increase the final expected concentration of the winner. This is likely due to having a different total concentration of chemicals in the system. Table~\ref{tab:wta} shows accuracy with different thresholds and chambers. Thresholds are extracted using the eval dataset (n=10000) and then tested on a separate dataset (n=10000). The "all" column represents one threshold used for all possible numbers of chambers, averaging the resulting accuracy. Figure~\ref{fig:binary_leader_d} shows a complete mapping of inputs to outputs for the case of two chambers. The Appendix~\ref{app:mcwta} shows the example run and the full description of the resulting system.

\section{Discussion and Limitations}
\label{sec:discussion}

We propose a new method of designing compact sparse Chemical Reaction Networks for solving a variety of computational problems. Previous work on CRN design focuses on construction of chemical counterparts of traditional basic computational units, such as logic gates, and manually combining them into circuits. In contrast, we show that end-to-end differentiable optimization is a viable approach to the objective-driven synthesis of complete circuit. This may enable efficient design of reaction circuits that can be implemented on a variety of physical substrates, from molecular to community scales.

We see following limitations and future research directions for this work: (A) All networks described here operate in a bulk, deterministic setting. Low molecular counts make systems stochastic and noisy, which brings both new challenges and opportunities. (B) Our sparsity regularization method removes all low-rate reactions from the system, although such reactions may sometimes be necessary for efficient implementation of the target function.

\medskip

\bibliography{refs}

\newpage

\appendix

\section{Appendix: Structure and dynamics of waveform-matching CRNs}

\begin{figure}[h]
    \centering
    \includegraphics[width=0.49\textwidth]{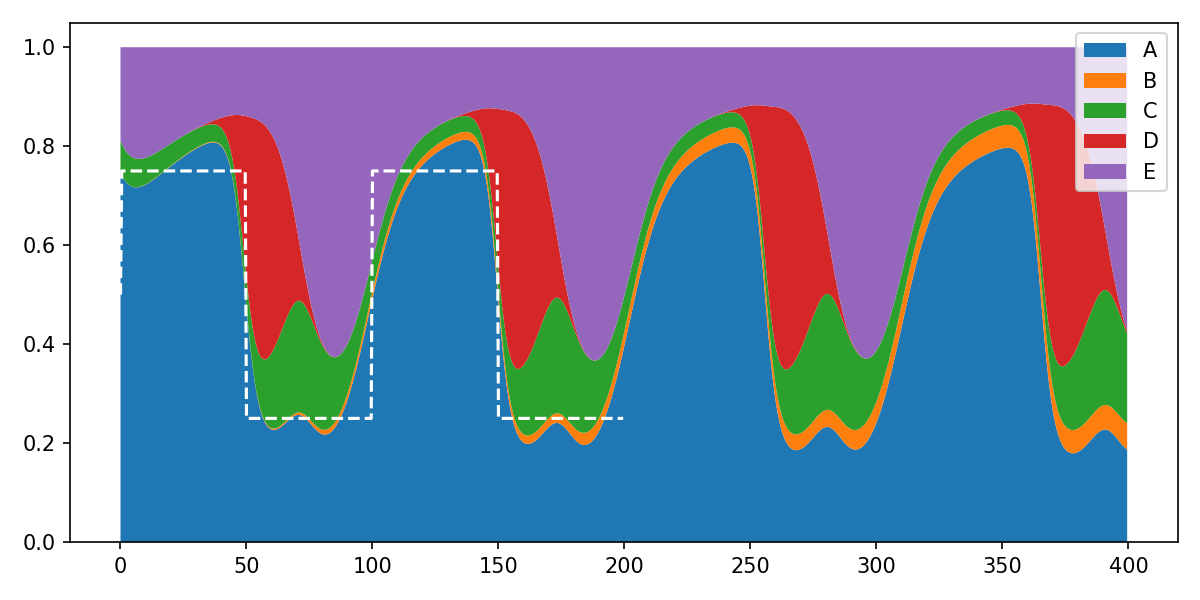}
    \includegraphics[width=0.49\textwidth]{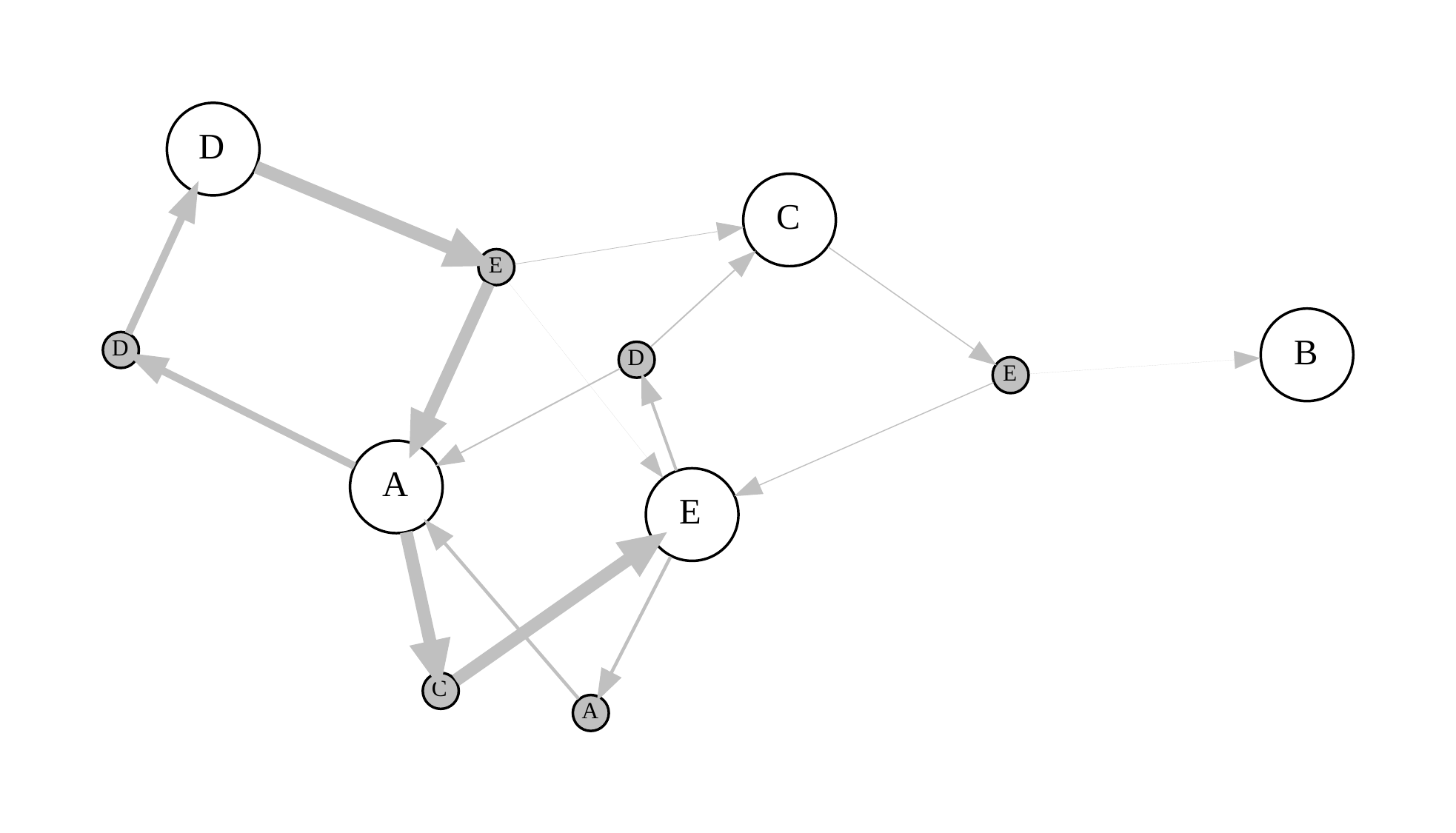}
    reaction count: 10 error: 5.661e-03 \\
    \includegraphics[width=0.49\textwidth]{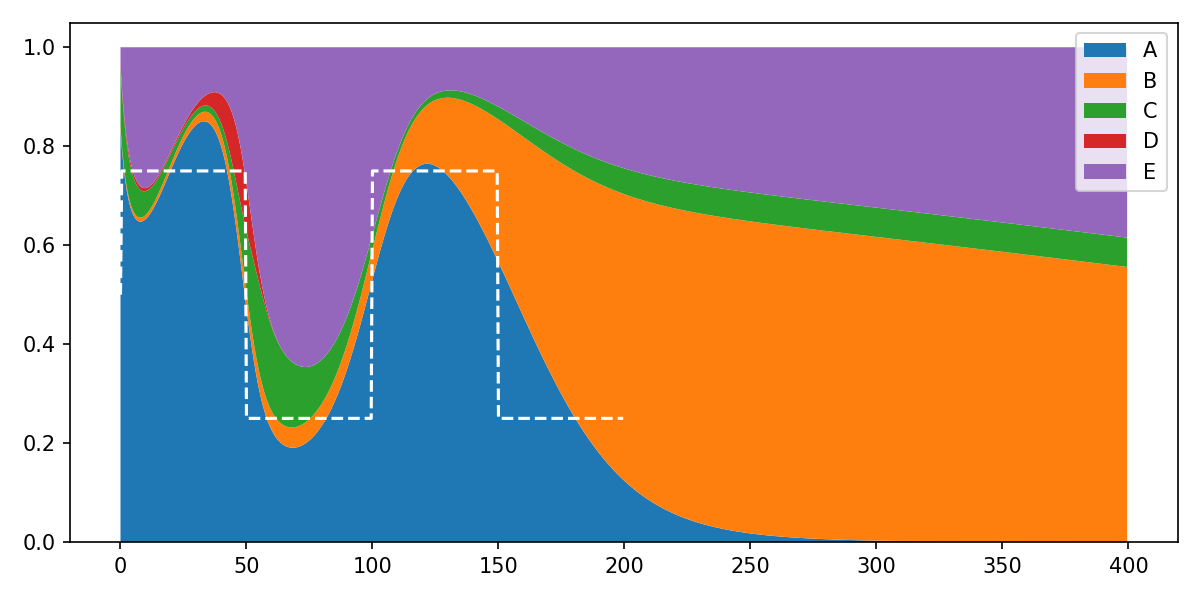}
    \includegraphics[width=0.49\textwidth]{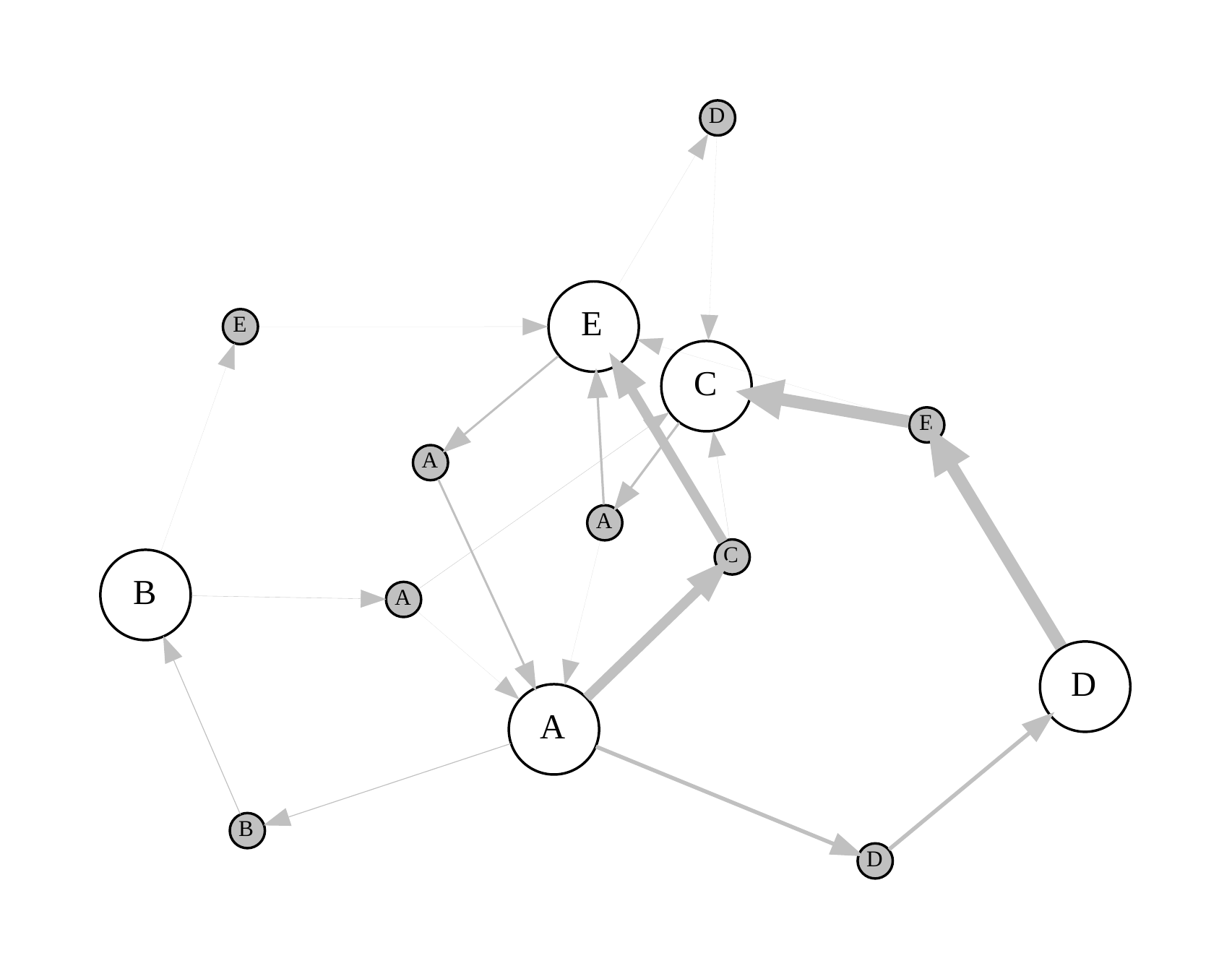}
    reaction count: 13 error: 1.244e-02 \\
    \includegraphics[width=0.49\textwidth]{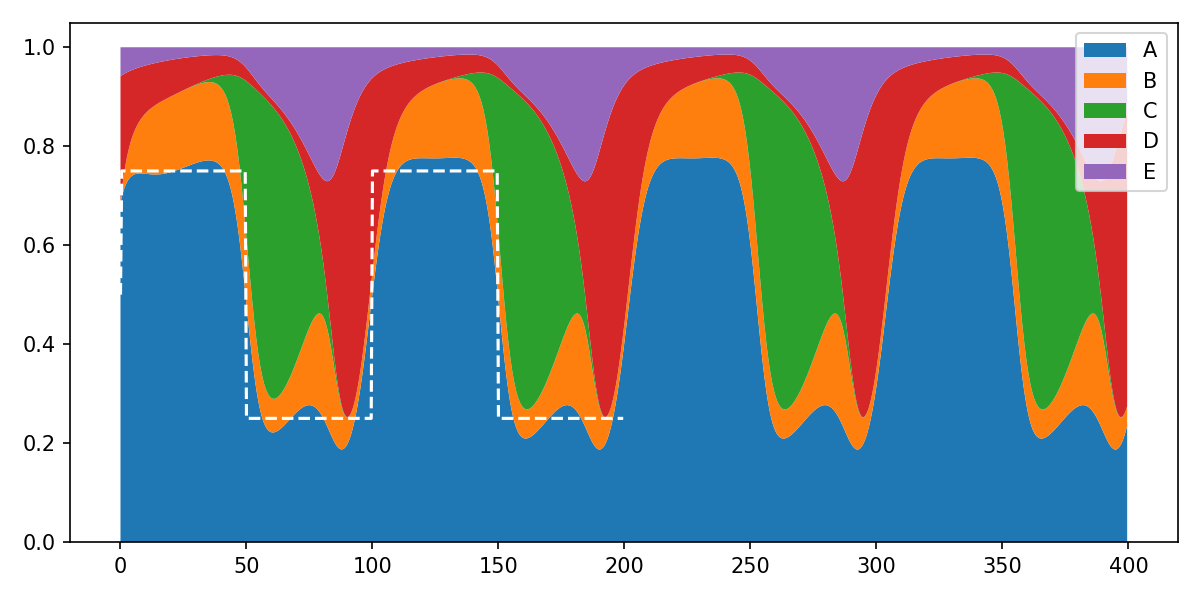}
    \includegraphics[width=0.49\textwidth]{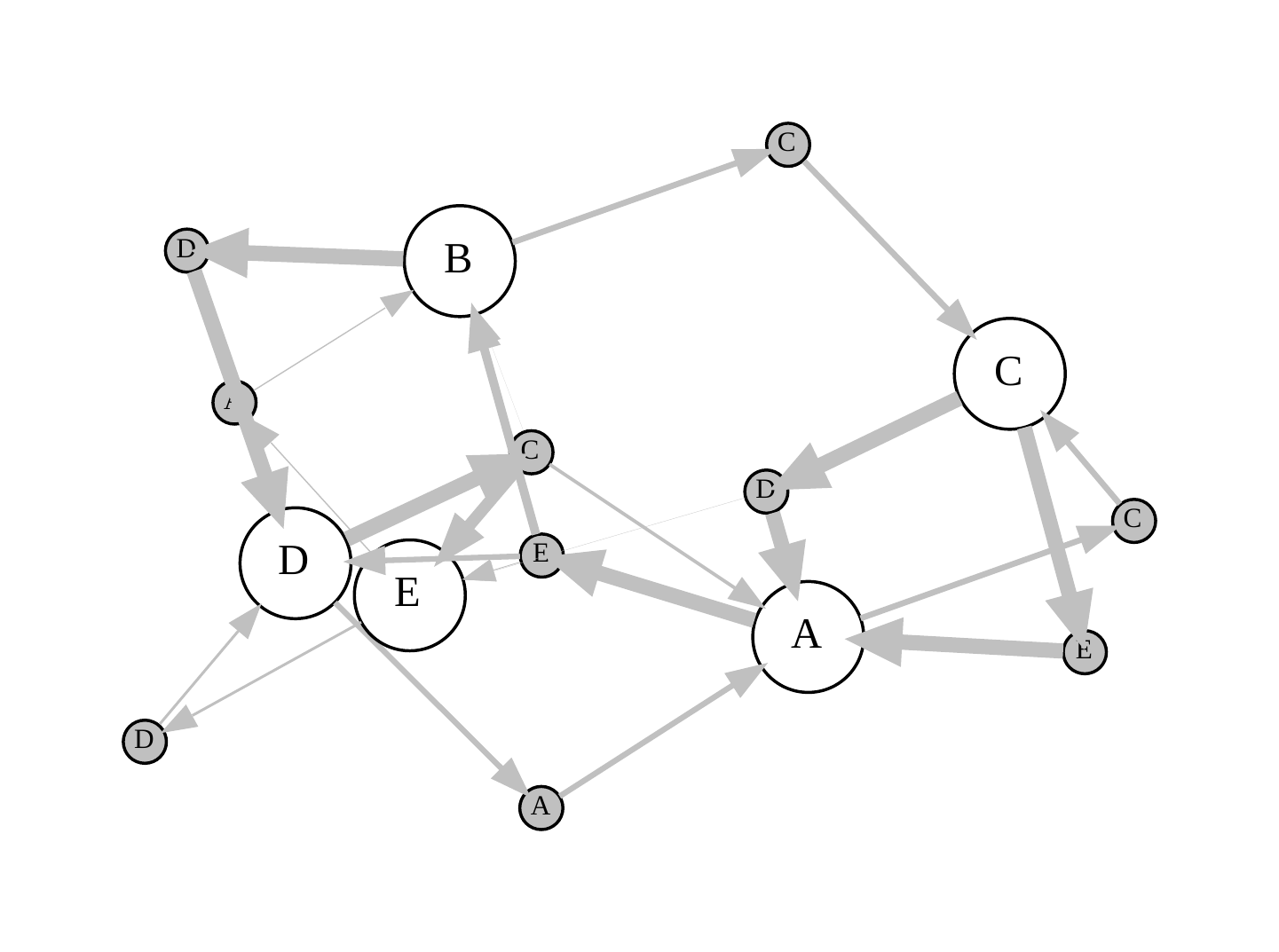}
    reaction count: 16 error: 4.313e-03 \\
    \caption{Examples of learned square oscillator network graphs (right) and their component dynamics (left). White nodes correspond to reactants and grey nodes to catalytic reactions. Edge width is proportional to the rate of the corresponding reaction.}
\end{figure}

\begin{figure}[h]
    \centering
    \includegraphics[width=0.49\textwidth]{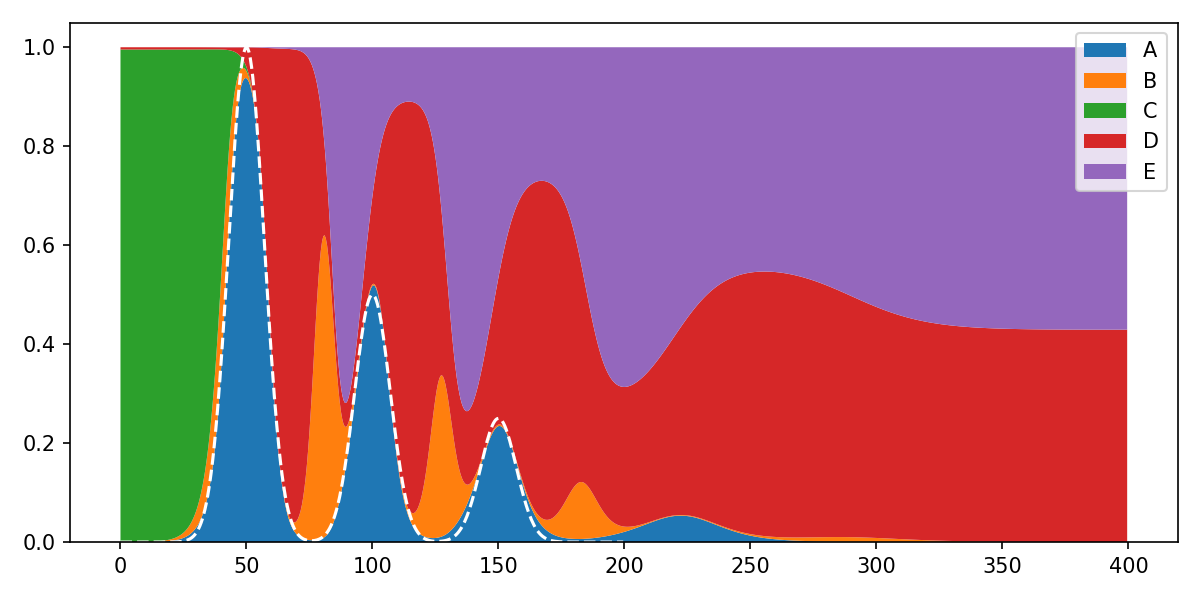}
    \includegraphics[width=0.49\textwidth]{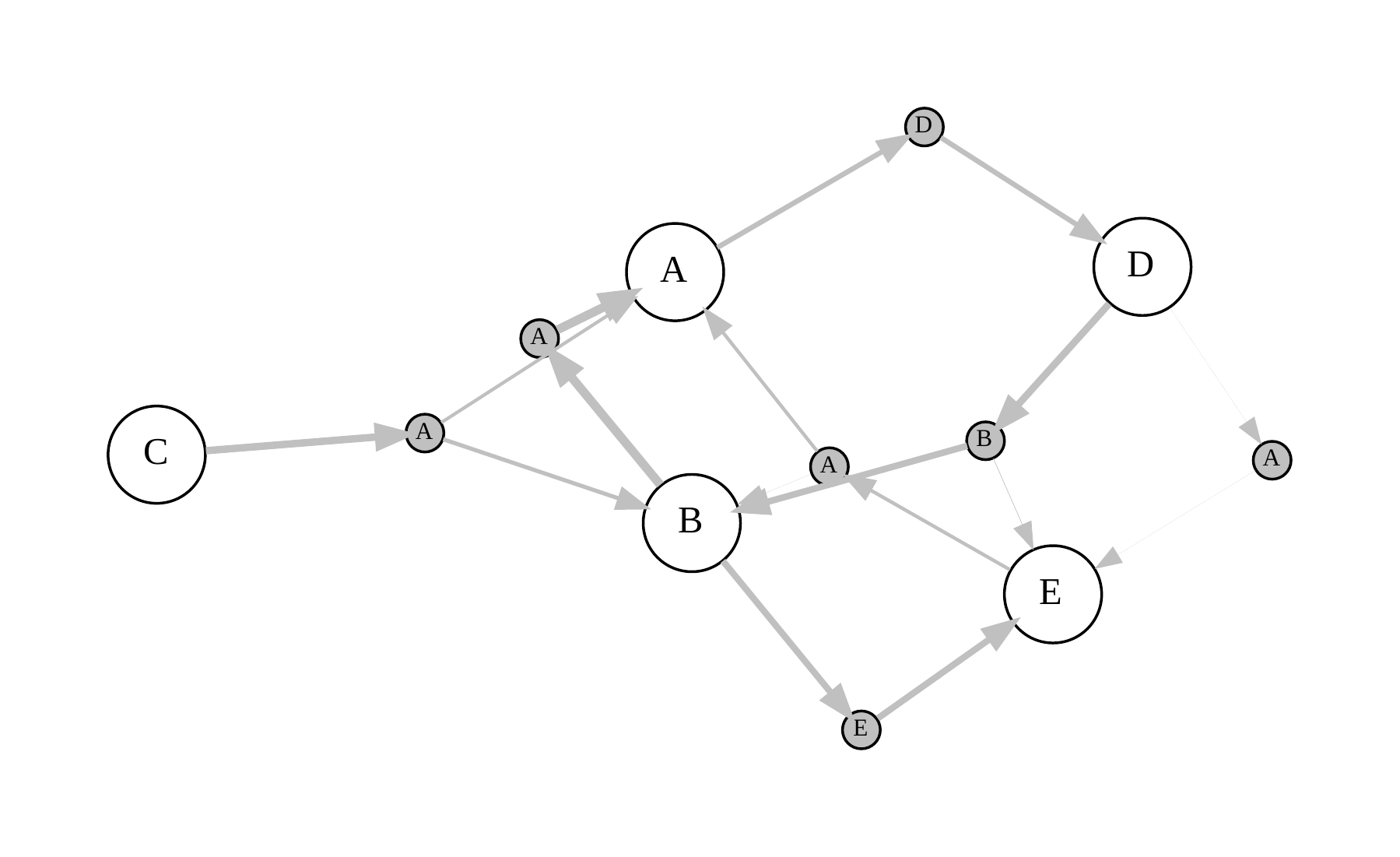}
    reaction count: 10 error: 2.390e-04 \\
    \includegraphics[width=0.49\textwidth]{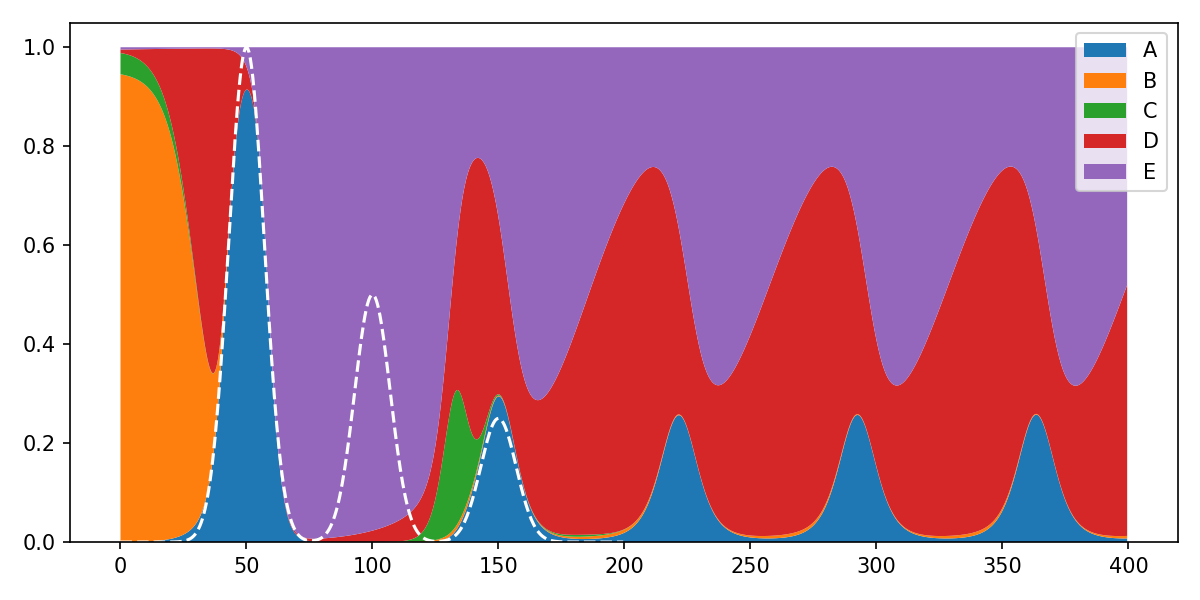}
    \includegraphics[width=0.49\textwidth]{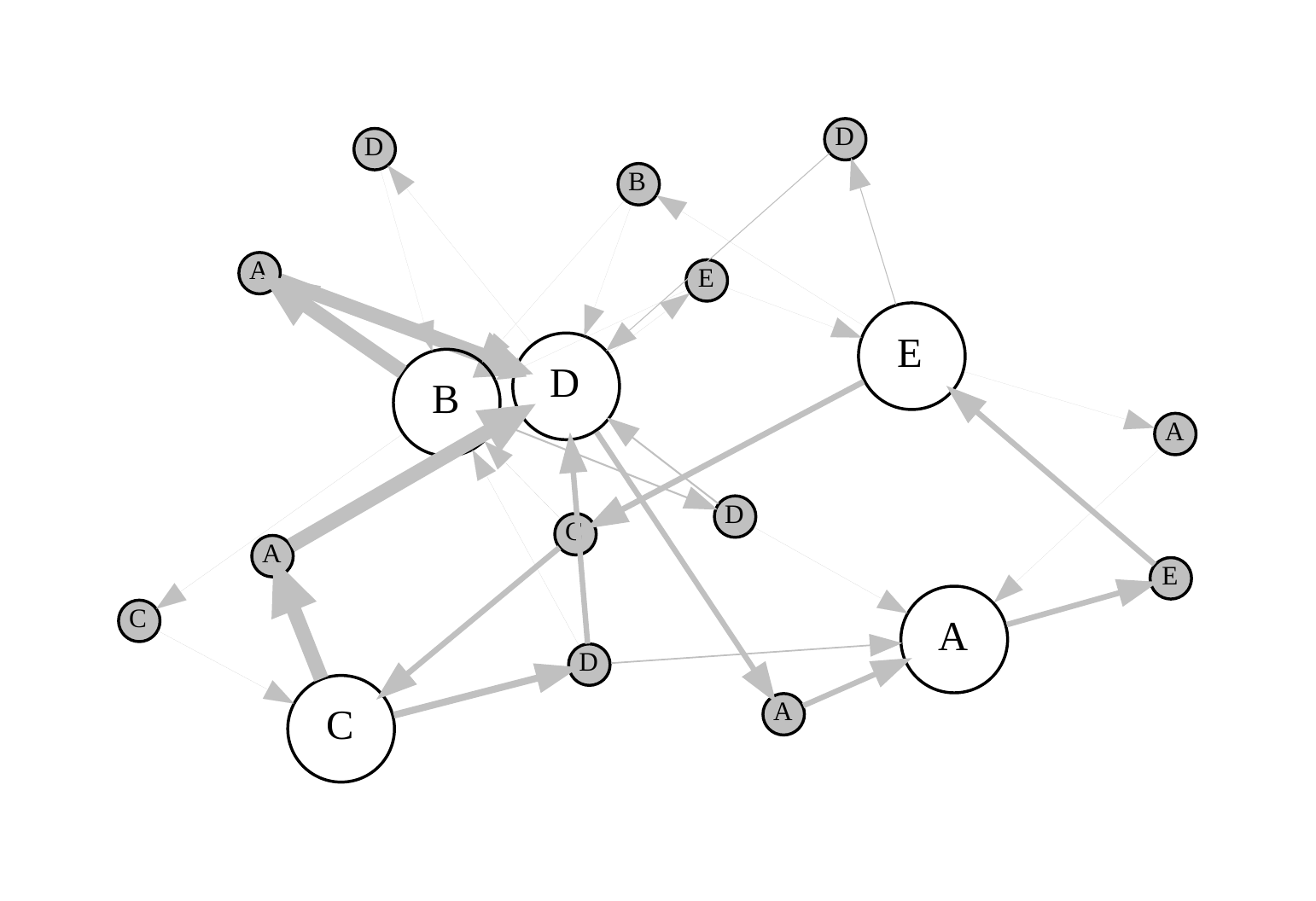}
    reaction count: 20 error: 1.603e-02 \\
    \includegraphics[width=0.49\textwidth]{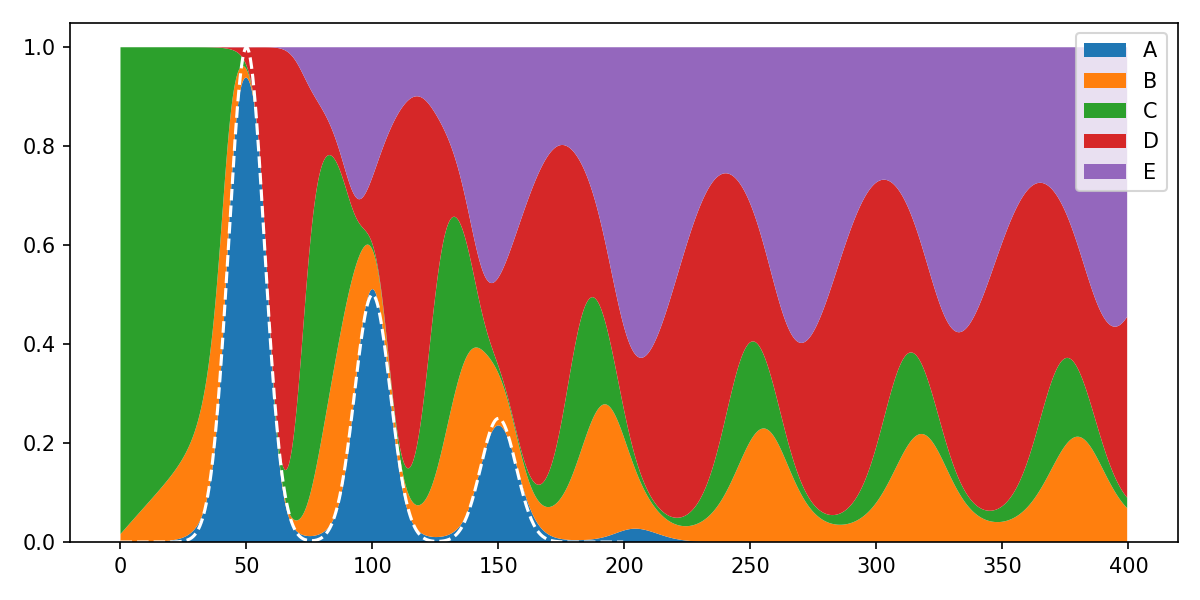}
    \includegraphics[width=0.49\textwidth]{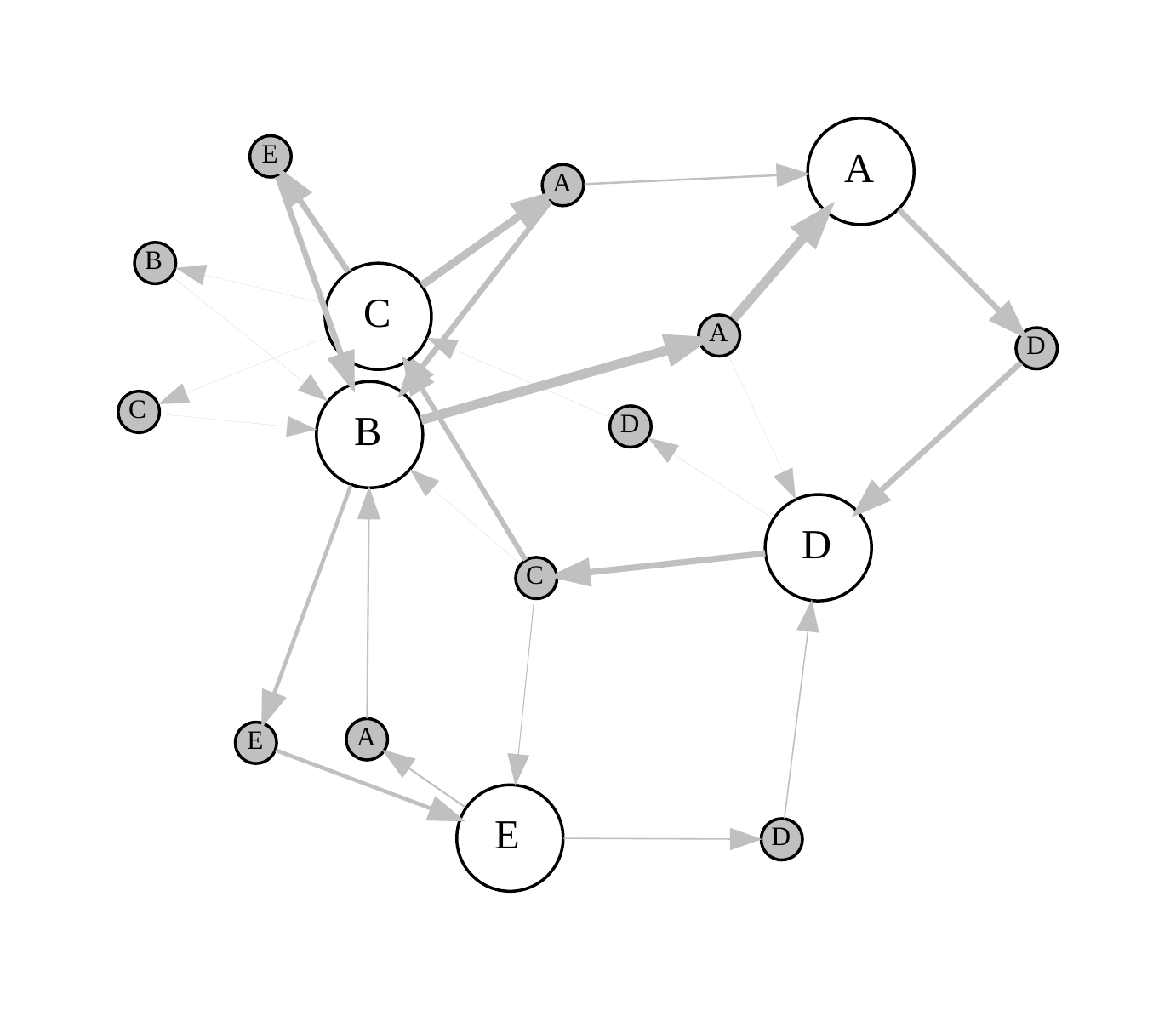}
    reaction count: 15 error: 1.803e-04 \\
    \caption{Examples of learned 'peaks' network graphs (right) and their component dynamics (left). White nodes correspond to reactants and grey nodes to catalytic reactions. Edge width is proportional to the rate of the corresponding reaction.}
\end{figure}

\clearpage

\subsection{Minimal differentiable CRN implementation in JAX}

\begin{lstlisting}[
language=Python,
caption=Minimal CRN ODE integration example. Running this code would produce the figure \ref{fig:ocs-plot},
xleftmargin=1.0cm]
import numpy as np
import jax
import jax.numpy as jp
import matplotlib.pylab as pl

def odeint_rk4(func, x0, t, *arg, **kw):
  '''Minimal RK4 ODE integrator (mimics scipy/jax odeint)'''
  def f(x, t):
    return func(x, t, *arg, **kw)
  def step_f(x, t_dt):
    t, dt = t_dt
    k1 = f(x, t)
    k2 = f(x+dt/2*k1, t+dt/2)
    k3 = f(x+dt/2*k2, t+dt/2)
    k4 = f(x+dt*k3, t+dt)
    x = x+(k1+2*(k2+k3)+k4)*(dt/6)
    return x, x
  t_dt = t[:-1], t[1:]-t[:-1]
  _, xs = jax.lax.scan(step_f, x0, t_dt)
  return jp.concatenate([x0[None], xs], 0)

def react(x, t_unused, T):
  '''CRN ODE derivative'''
  return jp.einsum('abc,...a,...b->...c', T, x, x)
  
T = np.zeros([3, 3, 3])
T[1, 0] = [-1, 1, 0]  # A -B-> B
T[2, 1] = [0, -1, 1]  # B -C-> C
T[0, 2] = [1, 0, -1]  # C -A-> A
x0 = jp.array([0.8, 0.1, 0.1])
t = jp.linspace(0, 30.0, 100)
y = odeint_rk4(react, x0, t, T)

pl.figure(figsize=(4.0, 2.5))
pl.plot(t, y);
pl.legend(['A', 'B', 'C'], loc='upper right')
pl.xlabel('time')
pl.ylabel('concentration')
\end{lstlisting}

\pagebreak
\section{Appendix: Additional Programs}

We present more examples of functions approximated by learned CRNs. For more training details please see the respective implementation in the notebook supplied in the supplementary materials.

\subsection{Analogue to Digital}

\begin{figure}
    \centering
 \begin{subfigure}[t]{0.17\textwidth}
     \centering
     \includegraphics[width=\textwidth]{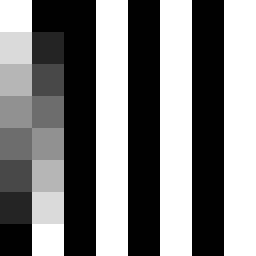}
     \caption{$t = 0$}
 \end{subfigure}
 \hfill
 \begin{subfigure}[t]{0.17\textwidth}
     \centering
     \includegraphics[width=\textwidth]{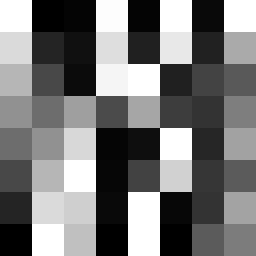}
     \caption{$t = 100$}
 \end{subfigure}
 \hfill
 \begin{subfigure}[t]{0.17\textwidth}
     \centering
     \includegraphics[width=\textwidth]{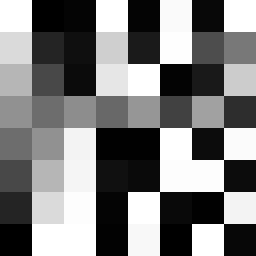}
     \caption{$t = 200$}
 \end{subfigure}
 \hfill
 \begin{subfigure}[t]{0.17\textwidth}
     \centering
     \includegraphics[width=\textwidth]{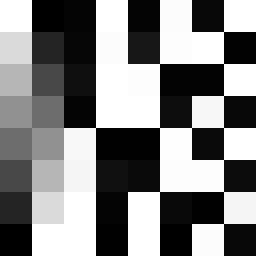}
     \caption{$t = 300$}
 \end{subfigure}
 \hfill
 \begin{subfigure}[t]{0.17\textwidth}
     \centering
     \includegraphics[width=\textwidth]{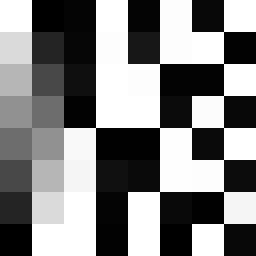}
     \caption{$t = 400$}
 \end{subfigure}
\caption{Visualisation of the evolution of a CRN that has been trained to map from analogue inputs (first and second columns), to a binary-encoded discretisation of the enumeration (row number) of the input (rightmost six columns of chemicals). Grayscale is used and chemical concentration $\in [0, 1]$ is linearly mapped to displayed intensity. We use dual-rail binary encoding with two chemicals "high" and "low" for each output variable. Each row denotes one set of chemicals being reacted according to the CRN. The CRN learns to correctly map the analogue-to-binary function. We don't display the values of three auxiliary chemicals that are used to aid the computation.}
    \label{fig:analogue_to_digital}
\end{figure}

We use the dual-rail encoding and learn a CRN capable of mapping from two analogue inputs $x \in [0,1]$ and complement $x_{c} = 1-x$, to a discretised binary representation of the enumeration of this value, using dual-rail encoding. We allow three auxiliary chemicals, and note that the CRN encodes this function correctly. See figure \ref{fig:analogue_to_digital} for a representation of the evolution of this CRN.

\subsection{Dynamically Controllable Oscillators}

\begin{figure}[h]
    \centering
    \includegraphics[width=0.49\textwidth]{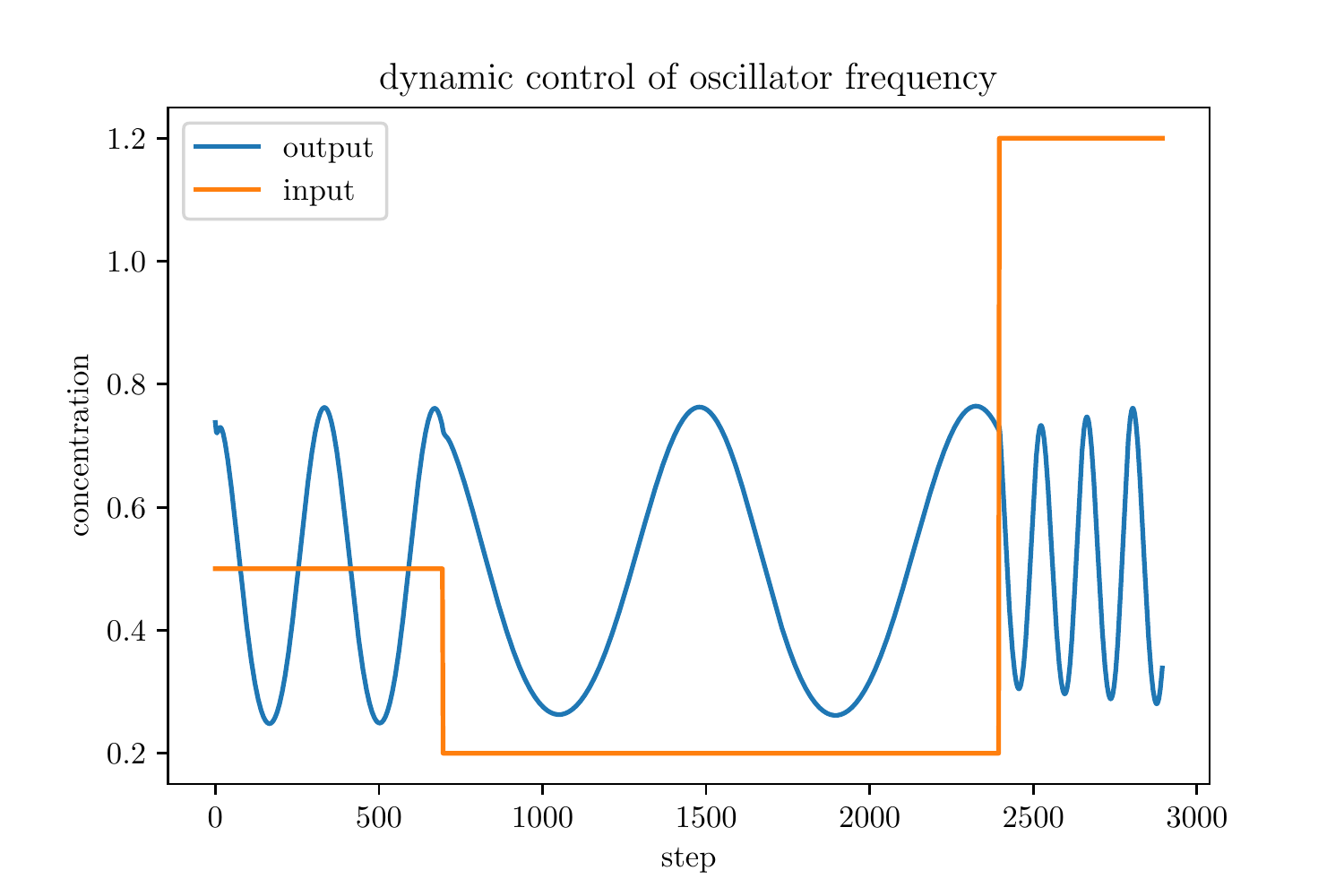}
    \includegraphics[width=0.49\textwidth]{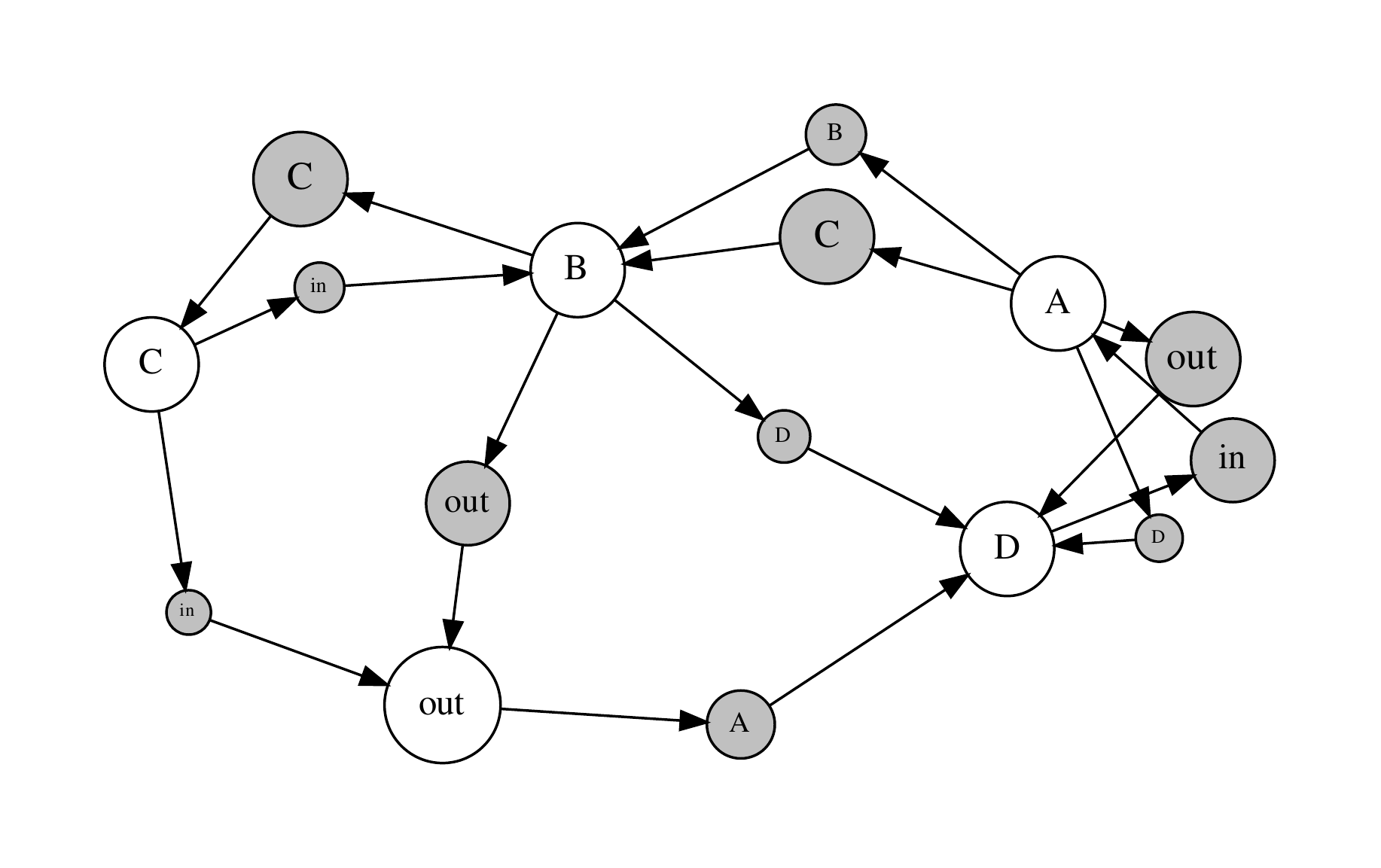}
    \caption{The dynamics and network graphs of an oscillator where the frequency of the oscillator is controlled by the second chemical ("input"). The "input" chemical is constrained to only act as a catalyst in the learned reactions. The levels of the "input" chemical can be seen on the dynamics graph. Note that changing the "input" while the CRN is oscillating results in a corresponding change in frequency of oscillation. The network graph displays reactions with a rate $>0.1$.}
    \label{fig:freq_osc}
\end{figure}

\begin{figure}[h]
    \centering
    \includegraphics[width=0.49\textwidth]{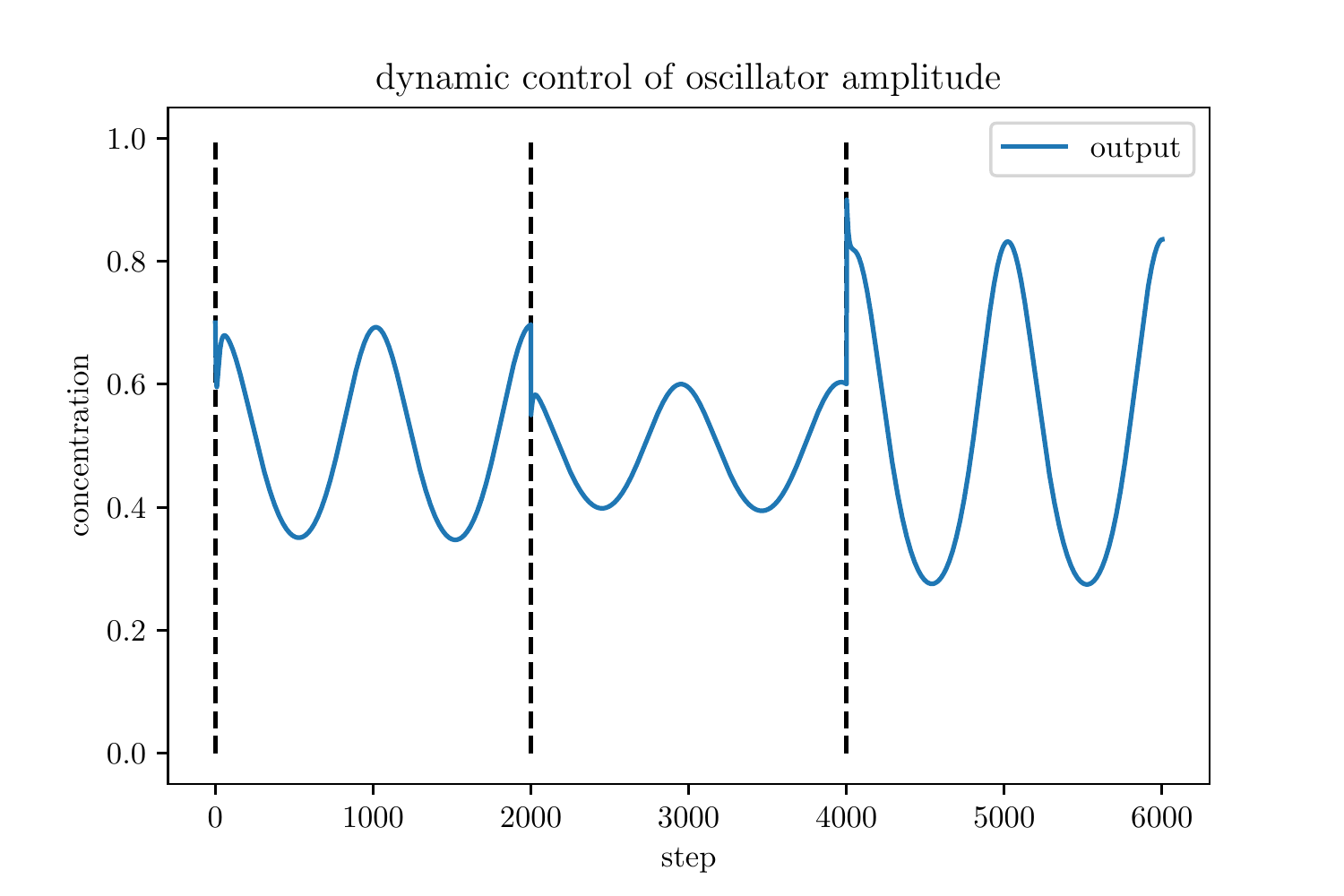}
    \includegraphics[width=0.49\textwidth]{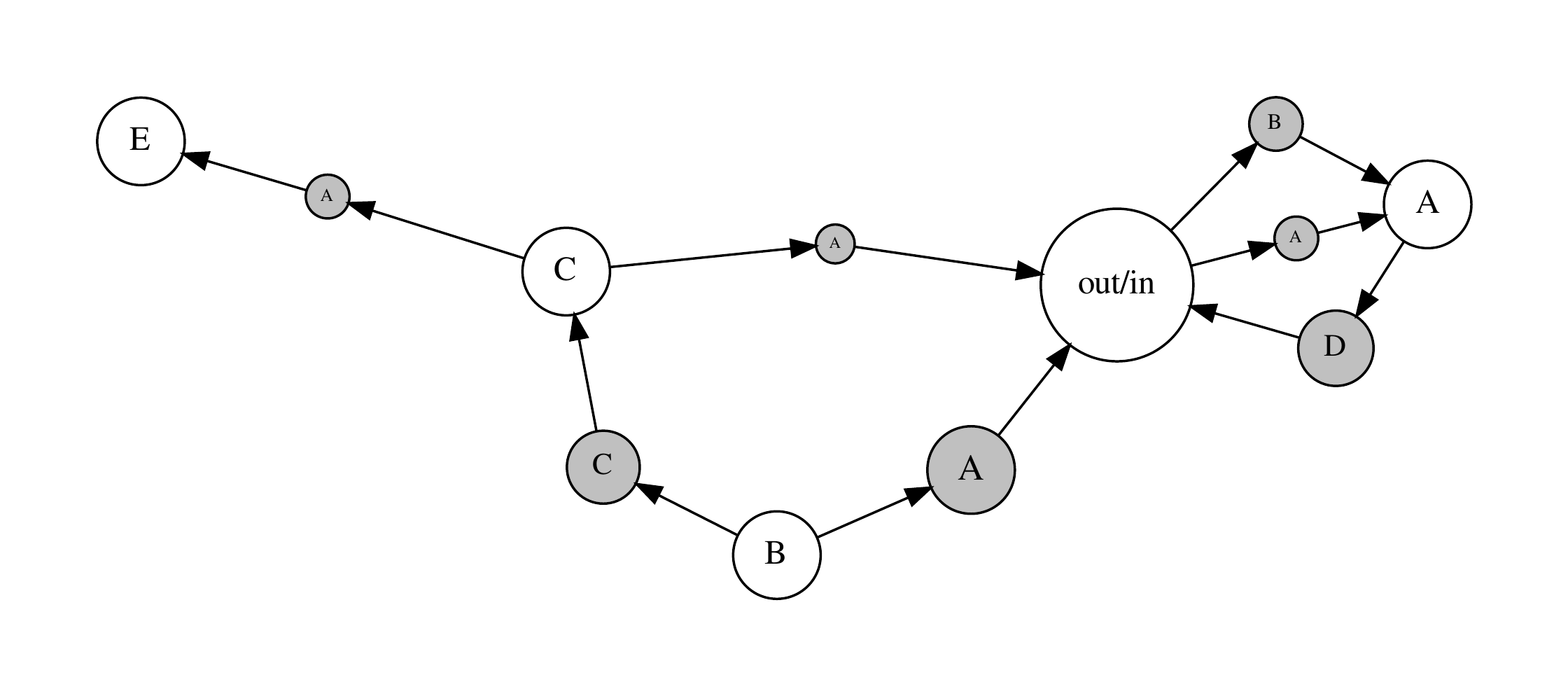}
    \caption{The dynamics and network graphs of an oscillator where the amplitude of the oscillator is determined by the initial concentration of the "output" chemical. The dashed lines indicate when the chemical of this chemical is dynamically changed during the reaction. A potential real life analogue to these events of increasing/decreasing a concentration would be adding more of input chemical to a mixture, or neutralising a certain amount of it. The network graph displays reactions with a rate $>0.1$.}
    \label{fig:amp_osc}
\end{figure}

We learn oscillators that are dynamically controllable using a "control chemical". We present two cases. The first is an oscillator where the frequency of oscillation is controlled by the concentration of the second chemical, and can be seen in \ref{fig:freq_osc}. The second is an oscillator where the amplitude of the oscillation in the first chemical matches the initial concentration of the first chemical, and can be seen in \ref{fig:amp_osc}. The two oscillator CRNs are trained using training batches containing a range of amplitudes and frequencies, respectively. See implementation for more details.

\section{Appendix: More details for the multi-chamber winner takes all task}\label{app:mcwta}

\subsection{Example run dynamics}

\begin{figure}
     \centering
     \begin{subfigure}[t]{0.31\textwidth}
         \centering
         \includegraphics[width=\textwidth]{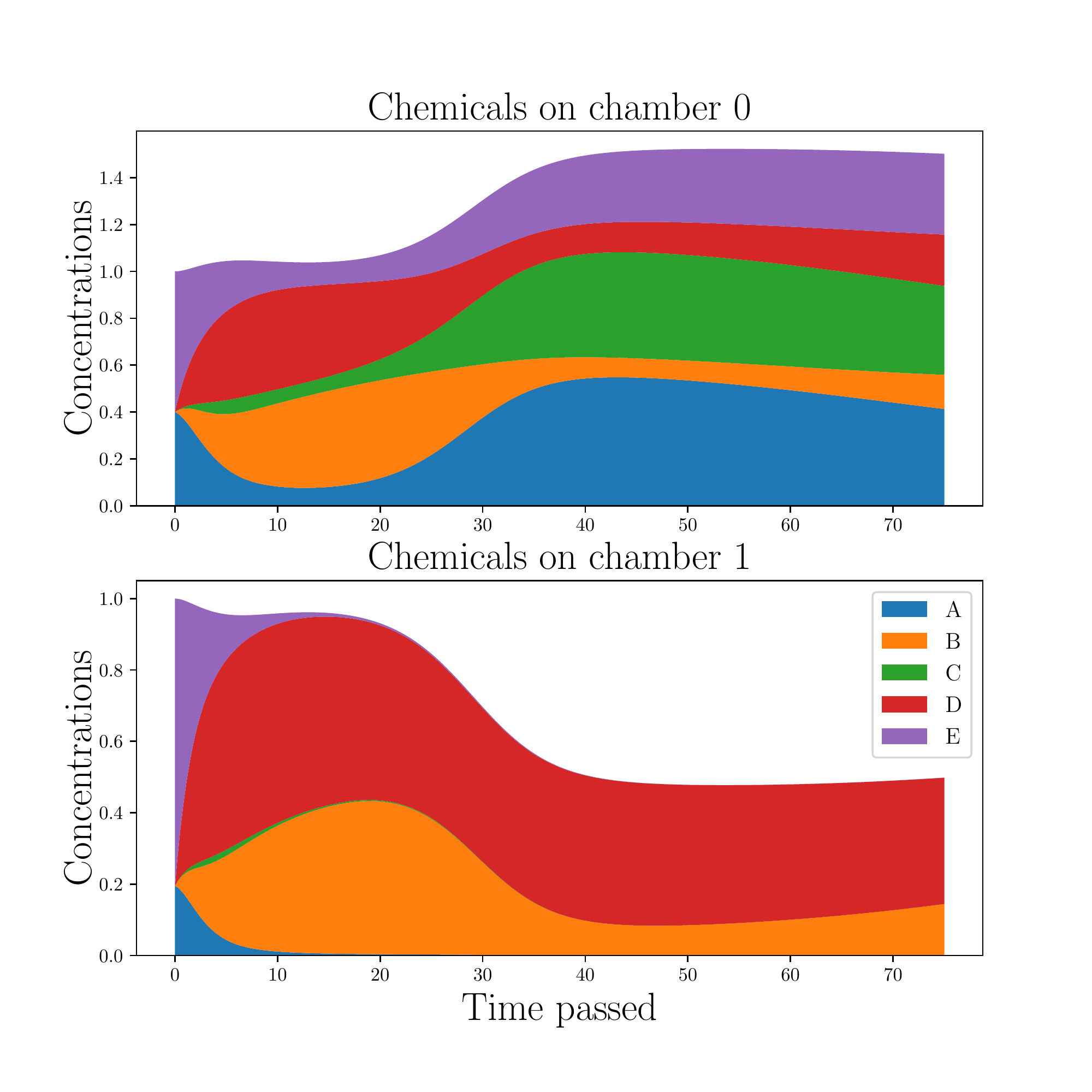}
         \caption{Chemicals dynamics (cumulative)}
         \label{fig:binary_leader_a}
     \end{subfigure}
     \begin{subfigure}[t]{0.31\textwidth}
         \centering
         \includegraphics[width=\textwidth]{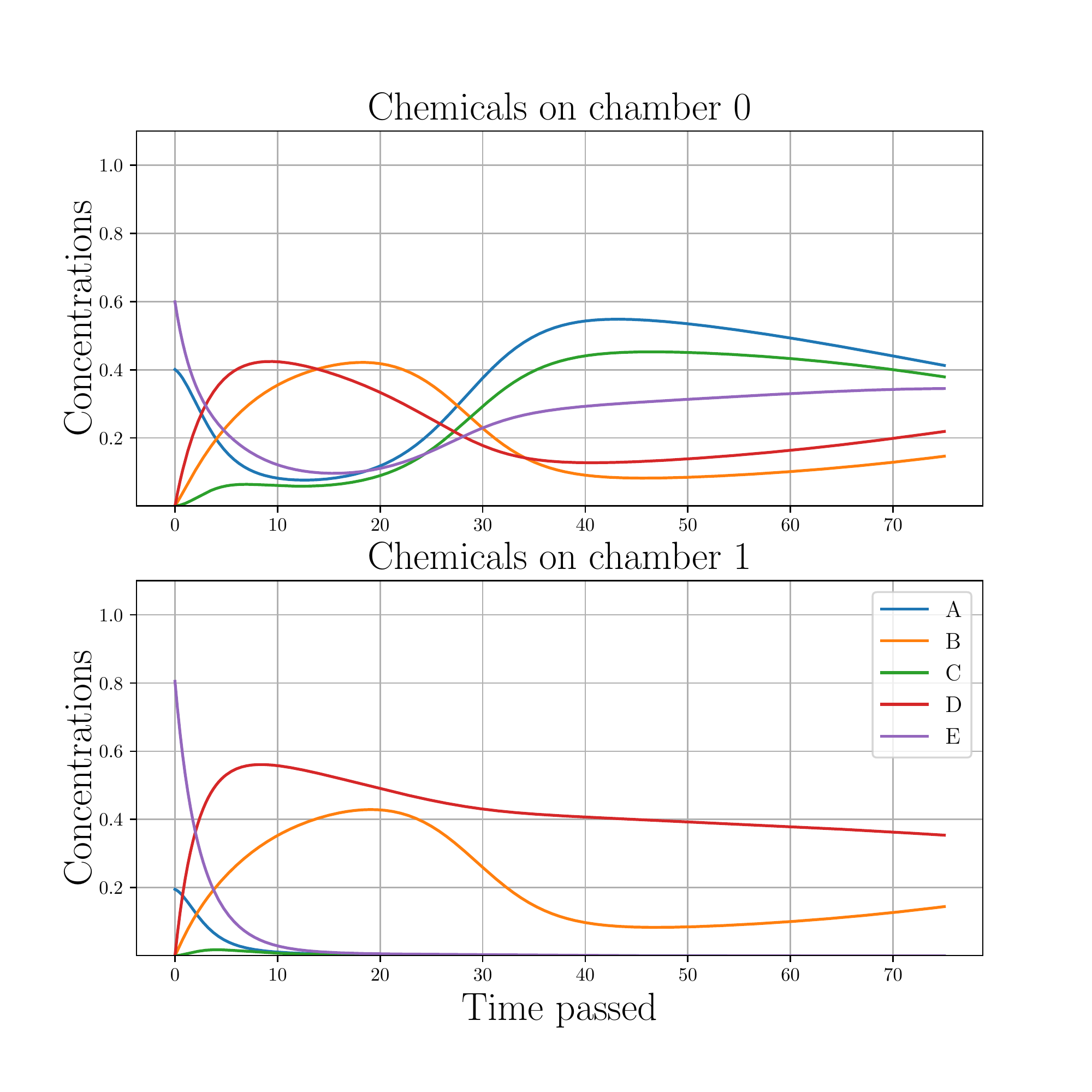}
         \caption{Chemicals dynamics (individual)}
         \label{fig:binary_leader_b}
     \end{subfigure}
     \begin{subfigure}[t]{0.31\textwidth}
         \centering
         \includegraphics[width=\textwidth, trim=0 -2cm 0 0]{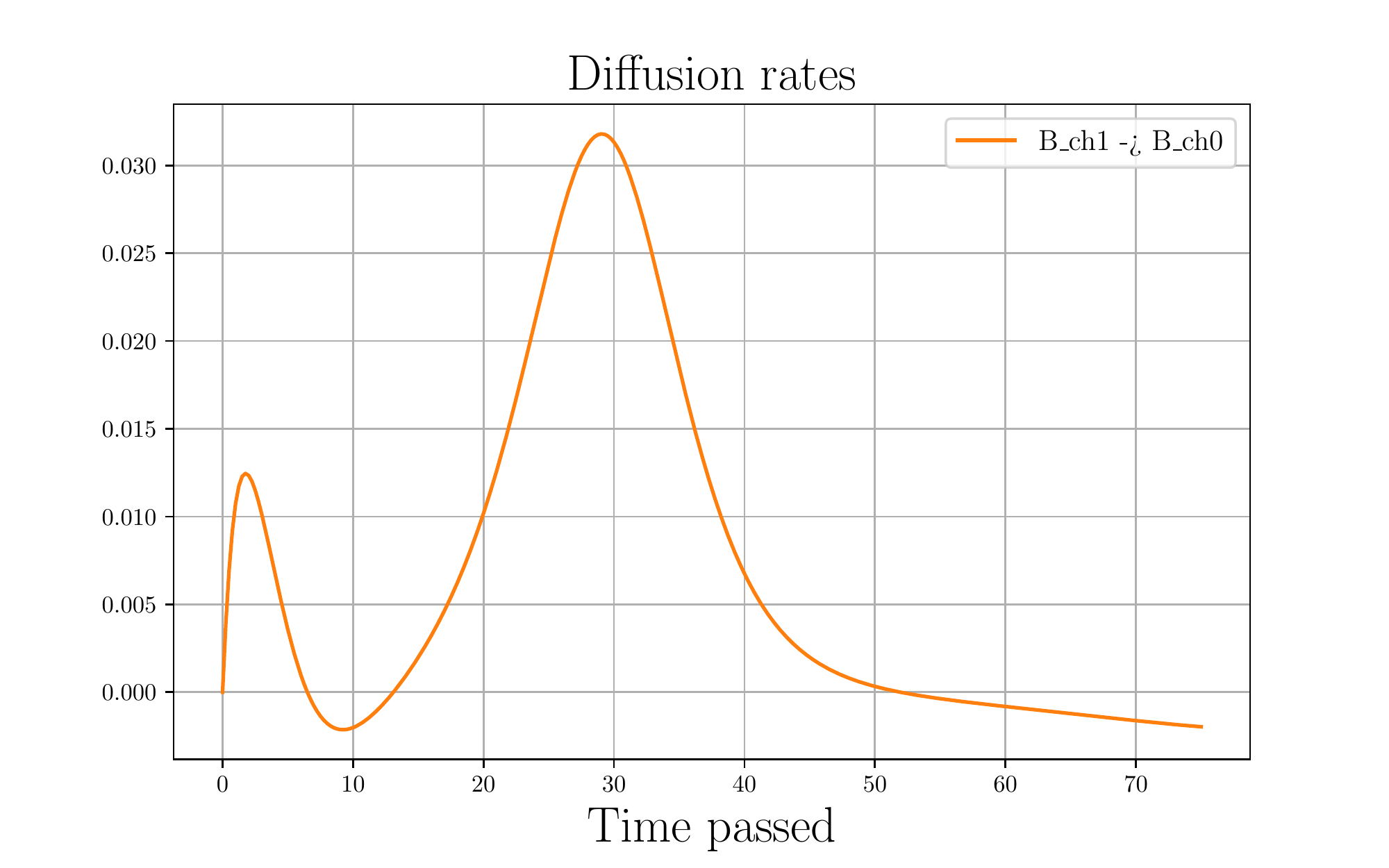}
         \caption{Diffusion}
         \label{fig:binary_leader_c}
     \end{subfigure}
\caption{Example run for binary winner selection dynamics. Plots (\subref{fig:binary_leader_a} and \subref{fig:binary_leader_b}) show time unfolding on chemical distributions for the two different chambers. Plot (\subref{fig:binary_leader_c}) shows the diffusion rates of the B chemical.}
    \label{fig:bls-dynamics}
\end{figure}

We can inspect the dynamics of chemicals for an example run of the two-chambers case in Figure~\ref{fig:bls-dynamics}. While it is beyond the scope of this paper to inspect the exact functioning of this network, we can observe a few properties. We note how the resulting model has only one chemical (chemical B) that diffuses in a significant way between chambers, while the other chemicals don't permeate (Figure~\ref{fig:binary_leader_c}), meaning that the relative difference in B is the signal used for leader selection. To understand where this relative difference in concentrations comes from, we can observe the values of the reaction tensor $T$ and note the highest possible value is assigned to $E \xrightarrow{1.0, B} B$, meaning that E gets converted to B as fast as possible in a self-reinforcing manner. Because E is an input that is the inverse of the input A, lower values of A create higher values of E, resulting in more B being generated on loser chambers. Therefore, B gets diffused to the winning chamber. In this system, the chemical D acts as a suppressant of A and C. We can confirm this by again observing the values of the reaction tensor $T$, and note the very high values of  $A \xrightarrow{0.98, D} E$, meaning that D acts as a catalyst to transform A into E, and the more composite ( $C \xrightarrow{0.32, D} B, C \xrightarrow{0.09, D} D, C \xrightarrow{0.56, D} E$), resulting in C being converted into B,D and E if D is present. D, instead, is suppressed by C ($D \xrightarrow{0.86, C} E$), and C is generated largely by the presence of B and A ($B \xrightarrow{0.21, A} C, A \xrightarrow{0.29, B} C$). Finally, A is generated thanks to C ($B \xrightarrow{0.7, C} A$). In summary, it appears that there is a suppression dynamic between D and the pair (A,C).

\subsection{Complete description of the system}

The resulting system has a diffusion coefficient of 1 for the chemical B. All the other coefficients are negligible.

The following is the complete list of reactions. Note that since we trained the network sparsely, the reactions that are not present in this list have no impact on the dynamics of the system.

$A~\xrightarrow{0.006, A}~C, 
B~\xrightarrow{0.794, A}~A, 
B~\xrightarrow{0.205, A}~C, 
C~\xrightarrow{0.001, A}~A, 
D~\xrightarrow{0.004, A}~A,
D~\xrightarrow{0.043, A}~B,$

$D~\xrightarrow{0.005, A}~C, 
D~\xrightarrow{0.109, A}~E, 
E~\xrightarrow{0.013, A}~A, 
E~\xrightarrow{0.068, A}~B, 
E~\xrightarrow{0.002, A}~C,
E~\xrightarrow{0.008, A}~D,$

$A~\xrightarrow{0.291, B}~C, 
C~\xrightarrow{0.002, B}~A, 
D~\xrightarrow{0.036, B}~B, 
E~\xrightarrow{1.000, B}~B,$

$A~\xrightarrow{0.001, C}~C, 
B~\xrightarrow{0.700, C}~A, 
B~\xrightarrow{0.297, C}~C, 
C~\xrightarrow{0.135, C}~A, 
D~\xrightarrow{0.001, C}~A, 
D~\xrightarrow{0.093, C}~B,$

$D~\xrightarrow{0.859, C}~E, 
E~\xrightarrow{0.023, C}~A, 
E~\xrightarrow{0.396, C}~C, 
E~\xrightarrow{0.005, C}~D,$

$A~\xrightarrow{0.012, D}~B, 
A~\xrightarrow{0.002, D}~C, 
A~\xrightarrow{0.004, D}~D, 
A~\xrightarrow{0.980, D}~E, 
B~\xrightarrow{0.002, D}~D, 
C~\xrightarrow{0.319, D}~B,$

$C~\xrightarrow{0.086, D}~D, 
C~\xrightarrow{0.560, D}~E, 
E~\xrightarrow{0.118, D}~B, 
E~\xrightarrow{0.467, D}~D,$

$A~\xrightarrow{0.008, E}~B, 
A~\xrightarrow{0.003, E}~C, 
A~\xrightarrow{0.015, E}~D, 
A~\xrightarrow{0.066, E}~E, 
C~\xrightarrow{0.034, E}~A, 
C~\xrightarrow{0.009, E}~B,$

$C~\xrightarrow{0.002, E}~D, 
D~\xrightarrow{0.002, E}~B, 
E~\xrightarrow{0.001, E}~A, 
E~\xrightarrow{0.082, E}~B, 
E~\xrightarrow{0.002, E}~C, 
E~\xrightarrow{0.439, E}~D$
\end{document}